\documentclass{llncs}
\pdfoutput=1
\usepackage{amsfonts,amsmath}

\usepackage[bookmarks=false]{hyperref}

\usepackage{alltt} 
\usepackage{dsfont} 
\usepackage{xspace}
\usepackage{tikz}
\usetikzlibrary{calc,matrix,arrows}
\usetikzlibrary{shapes,arrows,calc}
\usetikzlibrary{positioning}
\usetikzlibrary{trees, mindmap}
\usepgflibrary{arrows}

\usepackage{mathdots}

\usepackage{dsfont}
\usepackage{subfig}

\newcommand{\bsr}{bsr\xspace}
\newcommand{\bir}{bir\xspace}

\newcommand{\mbsr}{\mathbf{bsr}}
\newcommand{\mbir}{\mathbf{bir}}

\newcommand{\dform}[2]{[{#1}]{#2}}

\newcommand{\CTwoDat}{$\mathrm{C^{2}}+\mathrm{{Datalog}}$\xspace}

\newcommand{\CrtDatBsrBir}{$\mathrm{C_{\mathrm{r2}}^{2}}+\mathrm{Datalog+\{\mbsr,\mbir\}}$\xspace}

\newcommand{\BSTwo}{$\forall\forall$\xspace}
\newcommand{\FOTwo}{$\mathrm{FO^2}$\xspace}
\newcommand{\CTwo}{$\mathrm{C^2}$\xspace}

\newcommand{\NEXPTIME}{{\sc NExpTime}\xspace}
\newcommand{\NP}{{\sc NP}\xspace}

\newcommand{\EXPSPACE}{{\sc ExpSpace}\xspace}

\newcommand{\ProgP}{\mathbb{P}}
\newcommand{\Prog}[1]{\mathbb{{#1}}}

\newcommand{\ProgSeqP}{\mathds{P}}

\newcommand{\node}{\mathord{\mathit{node}}}

\newcommand{\lft}{\mathord{\texttt{left}}}
\newcommand{\rght}{\mathord{\texttt{right}}}
\newcommand{\NULL}{\mathord{\texttt{NULL}}}

\newcommand{\sem}[3]{\left({#1},{#2}\right)\leadsto{#3}}

\newcommand{\tuple}[1]{\langle#1\rangle}

\newcommand{\str}[1]{{\cal{#1}}}

\newcommand{\eq}{\mathbin{\approx}}
\newcommand{\noteq}{\mathbin{\not\approx}}

\newcommand{\notl}{\mathop{\neg}}
\newcommand{\andl}{\mathrel{\wedge}}

\newcommand{\iffl}{\mathrel{\Leftrightarrow}}
\newcommand{\redl}{\mathrel{\leftarrow}}

\newcommand{\Act}{\mathit{Act}}

\newcommand{\alloc}[1]{\mathit{alloc}({#1})}
\newcommand{\allocp}[1]{\mathit{alloc}'({#1})}

\newcommand{\gts}{\mathrel{:=}}
\newcommand{\assume}[1]{\mathtt{assume}({#1})}
\newcommand{\new}[1]{\mathtt{new}_{#1}()}
\newcommand{\dispose}[2]{\mathtt{free}_{#1}({#2})}

\newcommand{\nfield}{\texttt{next}}
\newcommand{\pfield}{\texttt{prev}}
\newcommand{\elt}{\texttt{c}}
\newcommand{\nelt}{\texttt{nc}}
\newcommand{\pelt}{\texttt{pc}}
\newcommand{\cdll}{\texttt{cl}}
\newcommand{\cdllnode}{\texttt{cl\_node}}

\newcommand{\ie}{i.\,e.,\xspace}

\newcommand{\eg}{e.\,g.,\xspace}

\newcommand{\wrt}{w.\,r.\,t.\xspace}

\newcommand{\resp}{respectively\xspace}
\newcommand{\etece}{etc.\xspace}

\begin{document}

\title{Bounded Model Checking of Pointer Programs
  Revisited\thanks{Research supported by NCN Grant no.  2011/03/B/ST6/00346}}

\author{Witold Charatonik\inst{1} and Piotr Witkowski\inst{2}}
\authorrunning{Witold Charatonik \and Piotr Witkowski}
\institute{Institute of Computer Science, University of Wroclaw,\\
\email{wch@cs.uni.wroc.pl}
\and
\email{pwit@pwit.info}}


\maketitle

\begin{abstract}Bounded model checking of pointer programs is
  a~debugging technique for programs that manipulate dynamically
  allocated pointer structures on the heap. It is based on the
  following four observations. First, error conditions like
  dereference of a dangling pointer, are expressible in a~fragment of
  first-order logic with two-variables. Second, the fragment is closed
  under weakest preconditions wrt.\ finite paths. Third, data
  structures like trees, lists etc.\ are expressible by inductive
  predicates defined in a fragment of Datalog. Finally, the
  combination of the two fragments of the two-variable logic and
  Datalog is decidable.

  In this paper we improve this technique by extending the
  expressivity of the underlying logics. In a~sequence of examples
  we demonstrate that the new logic is capable of modeling more
  sophisticated data structures with more complex dependencies on
  heaps and more complex analyses.
 \end{abstract}

\section{Introduction}
Automated verification of programs manipulating dynamically allocated
pointer structures is a challenging and important problem. 

In \cite{ChGM05} the authors proposed a bounded model checking (BMC) 
procedure for imperative programs that manipulate dynamically
allocated pointer structures on the heap. Although in this procedure
an explicit bound is assumed on the length of the program execution,
the size of the initial data structure is not bounded. Therefore, such
programs form infinite and infinitely branching transition
systems. The procedure is based on the following four
observations. First, error conditions like dereference of a dangling
pointer, are expressible in a~fragment of first-order logic with
two-variables. Second, the fragment is closed under weakest
preconditions wrt.\ finite paths. Third, data structures like trees,
lists (singly or doubly linked, or even circular) are expressible in a
fragment of monadic Datalog. Finally, the combination of the
two-variable fragment with Datalog is decidable. The bounded model
checking problem for pointer programs is then reduced to the
satisfiability of formulas in the combined logic. The authors gave an
algorithm solving the satisfiability problem in 2\NEXPTIME.

In this paper we further develop this method.  We formulate a general
BMC problem (which was not formulated earlier; it was only applied in
a rather ad hoc manner) and show that BMC of null pointers from
\cite{ChGM05} is its instance. We also provide several other instances
(Examples \ref{ex:aliasing}--\ref{ex:leaks}).  The logic used
in~\cite{ChGM05} was quite restrictive, in particular it allowed no
existential quantifiers and no sharing; it allowed only one Datalog
program; it could not express cardinality constraints; it had very
limited support for content analysis (only a single query to the
Datalog program was possible); it was unable to speak about inductive
(i.e., defined with Datalog) properties in postconditions.  Here, by
extending the expressivity of the underlying logics we are able to
model more sophisticated data structures with more complex
dependencies on heaps and more complex analyses.  Specifically, in a
series of examples we show the change in expressivity due to new
features of our logic: use of existential quantifiers (and other
relaxations to the syntax, like multiple queries to Datalog programs)
in Examples \ref{ex:halfShared}, \ref{ex:uml} and
\ref{ex:intersection}; sharing structure in Examples \ref{ex:lrlists},
\ref{ex:halfShared} and \ref{ex:intersection}; use of multiple Datalog
programs in Examples \ref{ex:intersection} and \ref{ex:leaks};
cardinality constraints in Example \ref{ex:halfShared}; support for
content analysis in Example \ref{ex:uml}; inductive properties in both
pre- and post-states in Example \ref{ex:intersection}.

It is worth noting that a bounded pointer program can traverse only a
bounded fragment of a data structure, which suggests that there is no
point in allowing data structures of unbounded size. However,
describing only traversable fragments is not enough for some analyses
that check certain properties of the whole heap. In Examples
\ref{ex:halfShared} (where we analyze cardinality constraints on the
heap as a whole) and \ref{ex:leaks} (where we detect memory leaks) the
result of the analysis depends on parts of the heap that are not
touched by the program.  These are properties of the heap as a whole,
and not of the traversable fragment.

\section{Related work}
There are many approaches to automated verification of pointer
programs that emerged recently. Most of them use logical formalisms to
describe heaps, capture program semantics and verify (partial) program
correctness by Hoare method. It requires expressing combinations of
heap shape with data properties, and quickly leads to undecidable
logics. Powerful proof engines (employing abstractions, theorem
proving and/or SMT based
reasoning)~\cite{ZeeKR09,RakamaricBHC07,LahiriQ08} are then used to
find proofs for specification obtained in that way allowing even the
\emph{full functional
  verification}~\cite{ZeeKR08,Leino10,Qiu0SM13,PekQM14} of data
structures, the holy grail of software analysis. Our approach is
different. We aim at bounded model checking, which allows to find bugs
rather than to prove their absence, and the logic we use is decidable,
what certainly limits its expressivity. A common belief is that pure
first-order logic is too weak to reason even about the simplest data
structures.  However the result from~\cite{WChPWit} implies that the
two-variable logic with counting \CTwo, a decidable first-order
fragment, is in some cases sufficient. There a combination of \CTwo
with Datalog was defined and shown to be reducible simply to
\CTwo. Here we continue this line of research by demonstrating
expressivity of another combination of \CTwo with Datalog that
translates directly to a decidable extension of \CTwo called \CTwo
with trees~\cite{ChW13,pwit2014}. Our longer term aim is to continue
investigation on expressive yet decidable extensions of first order
fragments applied to pointer program verification. Some specific
related work is discussed below.

\paragraph{Abstract interpretation and shape analysis.} One possibility is
to compute over-approximations of the set of reachable states and to
represent them, together with program actions, as formulas of a
3-valued logic with transitive closure. This is the approach taken \eg
in~\cite{SagivRepsWilhelm02,ImmermanRabinovichRepsSagivYorsh-CAV04,YorshRepsSagiv04,ManevichYRS05,ArnoldMSS06}. Its
soundness relies on abstract interpretation, but it may result in
false positives, \ie an erroneous state that is unreachable from any of
starting state may belong to the over approximation of the set of
reachable states. On the contrary, logics with Datalog are expressive
enough to precisely model reachable states of simple pointer
programs. However, these programs may as well be abstractions of other
pointer programs and abstract interpretation techniques might be 
applicable in our setting. A notable difference between our approach
and the one mentioned above is that a model of a formula with Datalog
represents concrete state of a program, while a model of 3-valued
formula represents (an abstraction of) a set of states. Recent work on
3-valued abstractions aims at verifying heap shape and data stored
there~\cite{FerraraFJ12,Ferrara14}. For simplicity, in our approach
only heap shape is represented, but since we represent heaps as models
of formulas that admit unary predicates, these can be used to simulate
finite domain data.

\paragraph{The Pointer Assertion Logic Engine.} Another option is to use
monadic second order logic on
trees~\cite{KlarlundSchwartzbach93,MollerSchwartzbach01}. Sets of
states of pointer programs are modeled using graph types, which
consist of tree backbones with some additional edges. As observed
in~\cite{ChGM05}, structures defined by Datalog programs can be seen
as tree backbones, but in our approach additional edges may be
specified in a fragment of first order logic, while graph types
specify these additional edges in a dynamic logic. The employed logic
is powerful, but of non-elementary complexity. In contrast, our
decidable logics with Datalog are relatively weak, but of an
elementary, \NEXPTIME\ complexity. This means that not all graph types
are expressible in our logics. On the other hand, due to arbitrary
binary predicates and to presence of cardinality constraints, our
logics are not subsumed by MSOL on trees.

\paragraph{Separation logic.} A powerful, but undecidable formalism for local
reasoning about pointer programs with lists was introduced
in~\cite{OHearnRY01}. This was a kind of proof-theoretical approach to
program verification, since proofs of Hoare triples must be manually
constructed using an intuitionistic proof calculus.  On the contrary,
both approaches discussed above as well as ours rely on decidable
logics. Decidable fragments of separation logics are also
studied~\cite{BerdineCalcagnoOHearn04,CalcagnoGH05,BrocheninDL08,BrocheninDL09,CookHOPW11,BansalBL09},
including fragments with not only lists but also general inductively
defined
predicates~\cite{NguyenDQC07,IosifRS13,AntonopoulosGHKO14,PiskacWZ14,BrotherstonFPG14}. Complexity
of these fragments vary from \NP\ through \EXPSPACE\ up to
nonelementariness. Our logic, \CTwoDat 
includes a
semantic restriction \bsr that, roughly, forces separation of
data structures defined by a single Datalog program. This suggests a
relation between our approach and the above mentioned. However, logics
with Datalog may also express structures that intersect, provided that
they are defined by different Datalog programs in a formula.

\paragraph{Logic for reachable patterns in linked data structures.} It
seems that in terms of expressibility the logic most related to ours
is the one from~\cite{YorshRabinovichSagivMeyerBouajjani}. It admits
arbitrary Boolean combinations of reachability constraints similar to
universally quantified guarded formulas. The exact difference in
expressive power needs to be investigated, but the two logics differ
in terms of complexity and underlying decision procedures.  The
satisfiability problem for the logic
in~\cite{YorshRabinovichSagivMeyerBouajjani} has \NEXPTIME lower bound
and elementary upper bound and it is also based on a translation to a
kind of monadic second-order logic on trees (the authors say that they
have another doubly-exponential procedure, but it is not
published). Although an arbitrary number of universally quantified
variables is allowed in the logic, the formulas in two examples
provided in~\cite{YorshRabinovichSagivMeyerBouajjani} use at most
three variables. Moreover, three variables are used only to define
properties that constitute semantic restrictions in our logic. We are
able specify all shapes occurring there in our combination of \CTwo
with Datalog. Moreover, although admitting only two variables, our
logic allows us unguarded quantification and counting quantifiers.

A significant difference between our approach to BMC of pointer
programs and many other BMC
techniques~\cite{BiereCCZ99,JacksonVaziri2000,ClarkeKroeningLerda-TACAS2004,HuthPradhan03}
is that we bound only the length of program paths to be symbolically
executed and not the size of the input data structures. Thus we are
able to perform model checking for infinite-state transition systems.

In addition to combined verification of heap shape with
data~\cite{FerraraFJ12,Ferrara14,PiskacWZ14}, heap shape with
size~\cite{NguyenDQC07} and balanceness of data
structures~\cite{HabermehlIV06} one may verify heap shape together
with content properties, where these properties are specified as
description logic formulas or UML
diagrams~\cite{CalvaneseKSVZ13,KotekSVZ14}. As we show in
Section~\ref{sec:expressive-power} our logic allows us to specify size
constraints as well as content properties, but not balanceness. Other
graph logics embedable in \CTwo for modeling heap shape as well as
content properties were earlier
considered~\cite{KuncakLR02,KuncakR04,Rensink04}. These logics are
incomparable with ours, since they aim at representing abstractions of
program states (a model represents a set of heaps), while logics with
Datalog are designed to represent concrete states (a model represents a
heap).

\section{Two-variable logic with counting and Datalog}\label{chap:prelim}
In this section we introduce our logic. It is a~two-variable fragment
of the first-order logic extended with counting quantifiers and
inductive predicates in form of Datalog programs. The obtained logic
is decidable by reduction to two-variable logic with counting and
trees~\cite{ChW13} and is expressive enough to model interesting
properties of dynamically allocated pointer structures.

\subsection{Monadic Datalog Programs}\label{sec:monadic-datalog}
Datalog is a declarative logic programming language. Syntactically it
is a subset of Prolog that does not use function symbols of arity
greater than $0$ (\ie constants). It is often used as a query language
for deductive databases. Here we use it to extend the expressive power
of some logics to be able to define dynamic structures on heap.
 
Let $\Sigma_E$ and $\Sigma_I$ be disjoint signatures, the former
(called \emph{extensional signature}) containing relational symbols of
arity at most 2, equality and constants, and the latter
(\emph{intensional signature}) containing only unary
symbols. Signature $\Sigma_{I}$ defines symbols that occur in heads of
clauses from a Datalog program, while remaining symbols occurring in
clauses come from $\Sigma_E$. We will call them \emph{intensional}
(\resp \emph{extensional}) symbols. Following~\cite{ChGM05}, we are
interested in \emph{monadic Datalog programs}. A clause in such a
program is a Horn clause where the only positive literal has a unary
predicate in its head and there are additional constraints on
remaining literals, as stated below.

\begin{definition}\label{program-Datalogowy}
  A~\emph{monadic Datalog program} over
  $\Sigma_{E}$ and $\Sigma_{I}$ is a finite set of clauses of~the~form
  \[p(u)\leftarrow B(u)\wedge
  \bigwedge_{i=1}^l[r_{i}\left(u,v_{i}\right)\wedge
  q_{i}(v_{i})],\text{ where}\] 
  \begin{enumerate}
  \item {$p(\cdot),q_{1}(\cdot),\ldots,q_{l}(\cdot)$ are
      $\Sigma_{I}$-predicates;}
  \item \label{program-Datalogowy:binarne}
    {$r_{1}(\cdot,\cdot),\ldots,r_{l}(\cdot,\cdot)$ are distinct
      $\Sigma_{E}$-predicates;}
  \item { $B(u)$,
      is a (possibly empty) quantifier-free first-order
      $\Sigma_{E}$-formula containing only constants and the
      variable $u$;}
  \item $l\geq0$ and $u,v_{1},\ldots,v_{l}$ are distinct variables.
  \end{enumerate}
  Monadic Datalog programs are further called Datalog programs for
  short. Datalog programs will be denoted by blackboard bold letters
  $\ProgP, \Prog{Q}, \Prog{R}$, $\ProgP_{\mathrm{list}}$ \etece
\end{definition}

\begin{figure}[h]
\begin{minipage}[c]{0.5\textwidth}
\begin{align*}
    \ProgP_{\mathrm{list}} = \{\; \mathrm{list}(x)&\leftarrow
    \mathrm{next}(x,y)\wedge \mathrm{list}(y),\;
    \\ \ \mathrm{list}(x)&\leftarrow x=\mathrm{NULL}\; \}.
  \end{align*}
 \end{minipage}
\begin{minipage}[c]{0.5\textwidth}
\newcommand{\nxt}{\mathord{\textit{next}}}

 \begin{tikzpicture}[wezel/.style={inner sep = 0.3mm, shape = circle,
      draw, fill}]

   \node (zero) [wezel, label=below:{\tiny $e_1$}] {};
   \node (jeden) [wezel, right of= zero,label=below:{\tiny $e_2$} ] {};
   \node (dwa) [wezel, right of=jeden, label=below:{\tiny $e_3$}] {};
   \node (null) [wezel,right of=dwa, label=below:{\tiny $\NULL$}] {} ;

\draw[-{latex}] (zero) -- node[above]{} (jeden);
\draw[-{latex}] (jeden) -- (dwa);
\draw[-{latex}] (dwa) -- (null);

\node (cjeden)[wezel] at (1.5,0.5) {};
\node (cdwa)[wezel] at (2,1) {};
\node (ccztery) [wezel]at (1,1) {};
\node (ctrzy) [wezel] at (1.5,1.5) {};

 \draw[-{latex}] (cjeden) to [bend right]  (cdwa);
 \draw[-{latex}] (cdwa) to [bend right] (ctrzy);
 \draw[-{latex}] (ctrzy) to [bend right] (ccztery);
 \draw[-{latex}] (ccztery) to [bend right] (cjeden);
\end{tikzpicture}
\end{minipage}
\caption {Datalog program $\ProgP_{\mathrm{list}}$ and a structure
  $\str M$. Edges represent the relation
  $\mathrm{next}(\cdot,\cdot)$.}\label{fig:simple}
\end{figure}
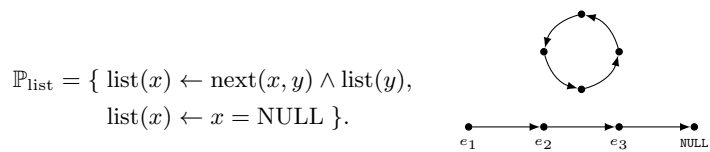

Consider an example Datalog program $\ProgP_{\mathrm{list}}$ from
Figure~\ref{fig:simple}. The extensional signature of
$\ProgP_{\mathrm{list}}$ is $\{\mathrm{next}(\cdot,\cdot), =,
\mathrm{NULL}\}$ and intensional signature is
$\{\mathrm{list}(\cdot)\}$. Our intention is that $\mathrm{list}(x)$
denotes that $x$ is a node of a singly linked list, where every node
is either $\NULL$ or has one successor pointed to by $\mathrm{next}$
pointer.

Datalog programs have natural least fixed point semantics. Given a
relational structure $\str M$ over $\Sigma_E$ and a Datalog program
$\ProgP$ over $\Sigma_E$ and $\Sigma_I$ the \emph{least extension of
  $\str M$ \wrt $\ProgP$} is the least $\Sigma_E\cup\Sigma_I$
structure $\str M_{\ProgP}$ such that 1) $\str M$ is contained in
$\str M_{\ProgP}$, and 2) if $[p(u)\leftarrow B(u)\wedge
  \bigwedge_{i=1}^l[r_{i}\left(u,v_{i}\right)\wedge q_{i}(v_{i})] \in
  \ProgP$ and $\str M_{\ProgP} \models B(e)
  \bigwedge_{i=1}^l[r_{i}\left(e,e_{i}\right)\wedge q_{i}(e_{i})]$
  then $\str M_{\ProgP}\models p(e)$, for all $e,e_1,\ldots,e_k \in
  \str M$. Consider the Datalog program $\ProgP_{\mathrm{list}}$ and
  structure $\str M$ from Figure~\ref{fig:simple}. The least extension
  of $\str M$ \wrt $\ProgP_{\mathrm{list}}$ is the structure $\str
  M_{\ProgP_{\mathrm{list}}} = \str M \cup \{\mathrm{list}(\NULL),
  \mathrm{list}(e_3), \mathrm{list}(e_2), \mathrm{list}(e_1)\}$.  The
  nodes in the cycle, are not members of the list; although the
  structure $\str M \cup \{\mathrm{list}(e)\mid e\in\str M\}$
  satisfies conditions 1) and 2) above, it is not the least one.

For a given Datalog program $\ProgP$ let $\Sigma(\ProgP)$ be the
subset of $\Sigma_E$ containing all binary predicates mentioned in
point \ref{program-Datalogowy:binarne} of Definition
\ref{program-Datalogowy}. 
Let
$\ProgP_1$, $\ProgP_2$ be Datalog programs over $\Sigma_E$ and
$\Sigma^1_I$, $\Sigma_E$ and $\Sigma^2_I$ respectively. Programs
$\ProgP_1$, $\ProgP_2$ are called \emph{disjoint} if $\Sigma^1_I \cap
\Sigma^2_I = \emptyset$. Let $\ProgP = \ProgP_1,\ldots,\ProgP_k$ be a
sequence of pairwise disjoint Datalog programs such that, for every
$i\in 1,\ldots,k$, the extensional vocabulary of $\ProgP_i$ is
$\Sigma_E$. The semantics of Datalog program sequence $\ProgP$ is the
same as semantics of a Datalog program $\bigcup_{i=1}^k \ProgP_i$: for
a structure $\str M$ over $\Sigma_E$ we define its least extension
\wrt sequence $\ProgP$ as $\str M_{\ProgP_1,\ldots,\ProgP_k} = \str
M_{\bigcup_{i=1}^k \ProgP_i}$.

\subsection{Syntax and semantics of the logic}\label{sec:ldat}

The logic we use in bounded model checking of pointer programs
combines the two variable logic with counting and inductive predicates
defined by Datalog programs. The two variable logic with counting
(\CTwo) is a~decidable fragment of first order logic containing
formulas whose all subformulas have at most two free variables, but
may contain counting quantifiers of the form $\exists^{\geq
  k},\exists^{=k},\exists^{\leq k}$. With these quantifiers one may
specify that there are at least, precisely or at most $k$ elements
with a given property.

We employ \CTwo formulas over
vocabulary $\Sigma_E \cup \Sigma_I$, but we impose some restrictions
on $\Sigma_I$ atoms that occur in these formulas. Let $\phi$ be a
$C^{2}$ formula over $\Sigma_{E}\cup\Sigma_{I}$ and let $\phi'$ be its
negational normal form. We say that a~$\Sigma_{I}$-atom $p(x)$ has
a~restricted occurrence in $\phi$ if either $p(x)$ occurs positively
in $\phi'$ and only in in scope of existential quantifiers or $p(x)$
occurs negatively in $\phi'$ and only in scope of universal
quantifiers.  For example $p(x)$ has a restricted occurrence in
formulas $\forall x\; p(x)\rightarrow\psi$, $\forall x\;(p(x)\wedge
q(x))\rightarrow\psi$, $\exists_{x}\ p(x)\wedge\psi$ or
$\exists_{x}\ p(x)\wedge q(x)\wedge\psi$, where $\psi$ is some $C^{2}$
formula with one free variable $x$ and no occurrence of $p(x)$,
and $q(\cdot)$ is some $\Sigma_{I}$-predicate. An occurrence of atom
$p(x)$ in formula $\forall_{y}\exists_{x}p(x)\wedge\psi$ is not
restricted, because $p(x)$ occurs positively and in scope of a
$\forall$ quantifier.

\begin{definition}[Syntax of \CTwoDat]\label{syntax} An expression 
  $\dform{\ProgP_1,\ldots,\ProgP_k}{\phi}$ is a \CTwoDat formula
 over $\Sigma_E$ and $\Sigma_I$ if
\begin{enumerate}
\item $\ProgP_1, \ldots, \ProgP_k$ are pairwise disjoint Datalog programs,
\item the extensional (\resp intensional) vocabulary of $\ProgP_i$ is
  contained in $\Sigma_E$ (\resp in $\Sigma_I$), for $i\in \{1,\ldots,
  k\}$,
\item $\phi$ is a formula of the two-variable logic with counting quantifiers
over the signature $\Sigma_{E}\cup\Sigma_{I}$, and 
\item every $\Sigma_{I}$-literal occurring in $\phi$ is an intensional
  literal defined by $\ProgP_1$ or $\ProgP_2$, or is a constant
  literal or has only restricted occurrences in
  $\phi$.\label{CrtDefinition}
\end{enumerate}
\end{definition}
Notice that Datalog programs $\ProgP_1$ and $\ProgP_2$ are
\emph{privileged}, \ie ($\Sigma_I^1 \cup \Sigma_I^2$)-predicates may
form arbitrary constant- or non-constant literals in $\phi$. On the
contrary, literals made of predicates defined by remaining Datalog
programs may either be constant literals or have only restricted
occurrences in $\phi$.

From now on, when a vocabulary $\Sigma_E$ is clear from context, we
will write $\mathrm{const}(v)$ as a shortcut for the formula
$\bigvee_{c\in \Sigma_E} v = c$. Let $\ProgP =
\ProgP_1,\ldots,\ProgP_k$ be a sequence of disjoint Datalog programs
such that $\ProgP_i$ is over $\Sigma_E$ and $\Sigma^i_I$, for $i\in
\{1,\ldots,k\}$. For a given $\Sigma_E$-structure $\str M$ let $\str
M_{\ProgP}$ be the least extension of $\str M$ \wrt sequence $\ProgP$.
We say that $\str M_{\ProgP}$ obeys the \emph{bounded-sharing}
restriction (\bsr for short) if $\str M_{\ProgP}$ is a model of all
sentences of the form \[\forall
u_{1},u_{2},v\;\left(s_{1}(u_{1},v)\wedge s_{1}(u_{2},v)\wedge
u_{1}\neq u_{2}\rightarrow \mathrm{const}(v)\right)\text{ and}\]
\[\forall
u_{1},u_{2},v\left(s_{1}(u_{1},v)\wedge
  s_{2}(u_{2},v)\rightarrow\mathrm{const}(v)\right),\] where $s_{1}$
and $s_{2}$ are two distinct predicates occurring in
$\Sigma(\ProgP_{i})$, for $i\in \{1,\ldots,k\}$.  We say that $\str
M_{\ProgP}$ obeys the \emph{bounded intersection} restriction (\bir
for short) if for all distinct predicates $p(\cdot),q(\cdot)\in
\Sigma^i_I$, structure $\str M_{\ProgP}$
models \[\forall{u}. p(u)\wedge q(u) \rightarrow \mathrm{const}(u).\]
Intuitively, the bounded-sharing restriction says that two pointers
occurring in the same Datalog program cannot point to the same memory
cell. The restriction ensures that data structures defined by a single
Datalog program are tree-like, in the sense that in-degree of their
nodes is $\leq 1$. Additionally, the bounded intersection restriction
forces these structures to be disjoint. In both cases an exception is
made for constant nodes; they can model \eg the $\NULL$ node which is
unique and shared among all data structures on heap, or the first
common node in two lists that have a common suffix.

\begin{definition}[Semantics of \CTwoDat]
  Let $\dform{\ProgP}{\phi}$ be a formula of \CTwoDat over
  $\Sigma_E$ and $\Sigma_I$, where $\ProgP = \ProgP_1,\ldots,\ProgP_k$
  and let $\str M_{\ProgP}$ be a finite structure over $\Sigma_{E}\cup
  \Sigma_I$ such that
  \begin{itemize}
  \item $\str M_{\ProgP}$ is the least extension of some
    $\Sigma_E$-structure $\str M$ \wrt Datalog program sequence
    $\ProgP$, 
  \item $\str M_{\ProgP}$ satisfies bounded-sharing and
    bounded-intersection restrictions, and
  \item $\str M_{\ProgP}\models\phi$.
  \end{itemize}
  Then $\str M_{\ProgP}$ is said to satisfy $\dform{\ProgP}{\phi}$,
  in symbols $\str M_{\ProgP}\models \dform{\ProgP}{\phi}$.
\end{definition}
Although both bounded-sharing and bounded-intersection restrictions
are expressible in our logics they cannot be removed, as they are
crucial in the satisfiability decision procedure
in~\cite{pwit2014}. With these restrictions we may express many data
structures including lists and trees, also with limited sharing of
substructures (see examples in the next section), but we cannot
express arbitrary DAGs.

Sometimes we would like to define linear constraints on the number of
realizations of unary predicates in a~structure $\str A$. When the
vocabulary $\Sigma$ is known from the context and $p_1(\cdot),\ldots,
p_l(\cdot)$ are unary symbols from $\Sigma$ we write $\Delta$ to
denote a system of linear (in)equalities in variables $\#p_1,\ldots,
\#p_k$. We say that $\str A$ satisfies $\Delta$ (written $\str A
\models \Delta$) if the valuation $\rho$ defined as $\rho(\#p_i) =
|p_i^{\str A}|$ satisfies $\Delta$. Here $|p_i^{\str A}|$ denotes the
number of elements of structure $\str{A}$ that satisfy the predicate
$p_i$. Let $\dform{\ProgP_1,\ldots,\ProgP_k}{\phi}$ be a \CTwoDat
formula and $\Delta$ be a system of linear (in)equalities over
intensional predicates from $\ProgP_1$ and $\ProgP_2$, and over unary
extensional predicates from $\Sigma_E$.  We write
$\dform{\ProgP_1,\ldots,\ProgP_k,\Delta}{\phi}$ for a formula with the
same semantics as the starting \CTwoDat formula, but with the
additional requirement that $\str M_{\ProgP_1,\ldots,\ProgP_k}\models
\Delta$.

The following theorem was proven in~\cite{pwit2014}, where \CTwoDat
was called \CrtDatBsrBir.

\begin{theorem}[\cite{pwit2014}, Cor. 3.27]
  Finite satisfiability problem for \CTwoDat, even enriched with linear
  (in)equalities, is \NEXPTIME-complete.
\end{theorem}
 The requirement that we allow at most 2 privileged Datalog programs
 in \CTwoDat formulas cannot be easily removed. It is related to an
 open problem, whether satisfiability for the two-variable
 logic \FOTwo with more than two successors of finite linear orders is
 decidable (note that we may express two successors of finite linear
 orders in \CTwoDat).

\subsection{Modeling data structures in  \CTwoDat}\label{sec:expressive-power}
Let us demonstrate expressive power of \CTwoDat.  We do it by writing
a handful of formulas describing heaps of imperative pointer
programs. Here we show examples of data structures, and in
Section~\ref{sec:analyses} we give examples of analyses that can be
modeled. More examples can be found in the PhD thesis of the second
author~\cite{pwit2014}.

A heap can be seen as a relational structure, where nodes are heap
elements (we assume that all elements of the heap are of the same
size), binary predicates denote pointers between nodes, constants
denote nodes pointed to by program variables and there is a
distinguished constant $\NULL$ denoting the null value. Binary
relations are interpreted as partial functions (\emph{functionality
restriction}) --- although every pointer on a heap has some value, in
our setting we allow it to have no value at all. Moreover, we
sometimes introduce auxiliary unary and binary predicates to express
additional properties. Note that the property of a binary predicate
$f(\cdot,\cdot)$ being a partial function is easily expressible in our
logic by a formula $\forall{x}\exists^{\leq 1}{y} f(x,y)$.

The logic \CTwoDat on structures that satisfy bounded-sharing and
bounded-intersection restrictions strictly subsumes the logics
considered in~\cite{ChGM05,WChPWit}. Therefore, after recalling the
simplest examples from~\cite{ChGM05}, we present exemplary structures
not expressible in the subsumed logics. Many other examples can be
found in~\cite{ChGM05,WChPWit,pwit2014}.  Let us start with a simple
example of a~singly-linked list.   
\begin{figure}
\newcommand{\nxt}{\mathord{\texttt{next}}}
\newcommand{\hconst}{\mathord{\texttt{h}}}
\centering
\subfloat[][$\phi = \mathrm{list(h)}$]{
\centering
  \begin{tikzpicture}[every node/.style={inner sep = 0mm, minimum size =
    0.45cm, shape = circle, draw}, thick, -latex, node distance = 35pt]
  
   \node (zero) {$e_0$};  
   \node (hhh) [ below of= zero, shape = rectangle] {$\hconst$};
   \node (jeden) [below of=hhh] {$e_1$};
   \node (null) [inner sep = 0.3mm, shape = rectangle, draw, below
     of=jeden] {$\NULL$} ;
     
  \node (dwa) [above of= null, left of= null] {$e_2$};
  \node (trzy) [above of= dwa, left of= dwa] {$e_3$};
  \node (cztery) [left of= dwa, below of= dwa] {$e_4$};
   
  \foreach \x in {zero,hhh,jeden,null, dwa, trzy}
       \node at (\x.east) [draw = none, right = 0pt] {\scriptsize
    $\mathrm{list}$};

    \node at (cztery.north) [draw = none, above = -3pt] {\scriptsize
    $\mathrm{list}$};

\draw (zero) -- node[right = 3pt, draw = none]{\scriptsize $\nxt$} (hhh) {};
\draw (hhh) -- node[right = 3pt, draw = none]{\scriptsize $\nxt$} (jeden) {};
\draw (jeden) -- node[right = 3pt, draw = none]{\scriptsize $\nxt$} (null) {};

\draw (dwa) -- node[sloped, above=-5pt, draw = none]{\scriptsize $\nxt$} (null) {};
\draw (trzy) -- node[above, sloped, above=-5pt, draw = none]{\scriptsize $\nxt$} (dwa) {};

\draw (cztery) -- node[auto, above =-5pt, draw = none]{\scriptsize $\nxt$} (null) {};
\end{tikzpicture}
 
 \label{fig:simple-list-left}
}
\qquad
\subfloat[][$\phi = \phi_1 \wedge \mathrm{list(h)}$]{
\centering
\begin{tikzpicture}[every node/.style={inner sep = 0mm, minimum size =
    0.45cm, shape = circle, draw}, thick, -latex, node distance = 35pt] 
  
   \node (hhh) [shape = rectangle] {$\hconst$};
   \node (jeden) [below of=hhh] {$e_1$};

     \node (null) [inner sep = 0.3mm, shape = rectangle, draw, below
     of=jeden] {$\NULL$} ;

  \foreach \x in {hhh,jeden,null}
       \node at (\x.east) [draw = none, right = 0pt] {\scriptsize
    $\mathrm{list}$};

 \draw (hhh) -- node[right = 3pt, draw =none]{\scriptsize $\nxt$} (jeden) {};
 \draw (jeden) -- node[right = 3pt, draw =none]{\scriptsize $\nxt$} (null) {};

\end{tikzpicture}
 \label{fig:simple-list-mid}
}
\qquad
\subfloat[][$\phi = \phi_1 \wedge \phi_2 \wedge \mathrm{list(h)}$]{
\centering

\newcommand{\prv}{\mathord{\texttt{prev}}}

  \begin{tikzpicture}[every node/.style={inner sep = 0mm, minimum size =
     0.45cm, shape = circle, draw}, thick, -latex, node distance = 35pt] 
  
   \node (hhh) [shape = rectangle, draw] {$\hconst$};
  
   \node (jeden) [below of=hhh] {$e_1$};

    \node (null) [inner sep = 0.3mm, shape = rectangle, draw, below
     of=jeden] {$\NULL$} ;

  \foreach \x in {hhh,jeden,null}
       \node at (\x.east) [draw = none, right = 0pt] {\scriptsize
    $\mathrm{list}$};

\draw (hhh) to [bend right] node[left= 3pt, draw = none]{\scriptsize $\nxt$} (jeden) {};
\draw (jeden) to [bend right] node[right =3pt, draw = none]{\scriptsize $\prv$} (hhh) {};
\draw (jeden) -- node[right = 3pt, draw = none]{\scriptsize $\nxt$} (null) {};

\end{tikzpicture}

 \label{fig:simple-list-right}
}
\caption{Models of $\dform{\ProgP_{list}}{\phi}$ from
  Example~\ref{ex:simple-list} for different formulas $\phi$. Here $\phi_1 =
  \left(\forall{u}\; \neg\mathrm{next}(u,h) \wedge
    \forall{u}\forall{v}\; \mathrm{next}(u,\NULL) \wedge
    \mathrm{next}(v,\NULL) \rightarrow u\eq v \right)$ and $\phi_2 =
  \forall{u}\forall{v}\; (u\neq \NULL \wedge v\neq \NULL) \rightarrow
  (\mathrm{next}(u,v)\leftrightarrow \mathrm{prev}(v,u))$.
}
\label{fig:simple-list}
\end{figure}
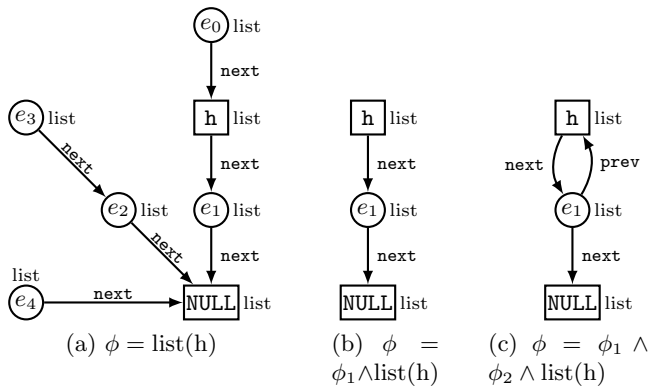

\begin{example}\label{ex:simple-list}
  The simplest linked data structure is a singly-linked
  $\NULL$-terminated list with head in some specified node
  $\mathrm{h}$. For $\Sigma_E = \{\mathrm{next}(\cdot,\cdot),
  \mathrm{h},\NULL\}$ and $\Sigma_I = \{\mathrm{list}(\cdot)\}$ let
  $\varphi = \dform{\ProgP_{\mathrm{list}}}{\phi}$ be a \CTwoDat
  formula where $\ProgP_{\mathrm{list}}$ is defined in 
  Figure~\ref{fig:simple}. By defining $\phi$ to be just a query
  $\mathrm{list(h)}$ we force models of $\varphi$ to contain a list
  from $h$ to $\NULL$ made of $\mathrm{next}$ edges. One of the
  possible models of $\dform{\ProgP_{\mathrm{list}}}{\phi}$ is
  depicted in Figure~\ref{fig:simple-list-left}. Thanks to
  bounded-sharing restriction $\NULL$ and $\mathrm{h}$ can be the only
  nodes shared by different lists --- in
  Fig.~\ref{fig:simple-list-left} node $\NULL$ is shared while
  $\mathrm{h}$ is not. Moreover, the functionality restriction ensures
  that every node emits at most one $\mathrm{next}(\cdot,\cdot)$
  pointer.  The bounded-intersection plays no role here, since there
  is only one intensional predicate in the signature.

We may ensure that $h$ is indeed the head of the list (and not an
internal node) by adding a conjunct $\forall{u}\;
\neg\mathrm{next}(u,h)$ to $\phi$. Moreover, to ensure that the list
with head in $h$ is the only list in the structure, we add to $\phi$
formula $\exists^{\leq 1}u\;\mathrm{next}(u,\NULL)$. This is depicted
in Figure~\ref{fig:simple-list-mid}. We can further modify our formula
to capture doubly linked lists by adding to $\phi$ a conjunct
$\forall{u}\forall{v}\; (u\neq \NULL \wedge v\neq \NULL) \rightarrow
(\mathrm{next}(u,v)\leftrightarrow \mathrm{prev}(v,u))$ as in
Figure~\ref{fig:simple-list-right}.
\end{example}
The $\NULL$ node represents the undefined memory address. It may be
pointed to by an arbitrary number of pointers, but no pointer can
start in it. This is expressed by a formula $\forall{u}\; \neg
\mathrm{next}(\NULL,u)$ in the context of the example above, and in
general by $\bigwedge_{r(\cdot,\cdot)\in \Sigma_E}\forall{u}\; \neg
r(\NULL,u)$. We assume that such a conjunct is implicitly included in
every formula we write here.

The next example shows the difference between modeling a~data
structure with a~single Datalog program and a~sequence of Datalog
programs.

\begin{example}\label{ex:lrlists}
 Consider the following Datalog programs.
  \begin{align*}
    \ProgP_{llist} = \{\; \mathrm{llist}(x)&\leftarrow \mathrm{left}(x,y)\wedge
    \mathrm{llist}(y),\; 
    \mathrm{llist}(x) \leftarrow x=\mathrm{NULL}\; \}. \\ \ 
    \ProgP_{rlist} = \{\; \mathrm{rlist}(x)&\leftarrow \mathrm{right}(x,y)\wedge
    \mathrm{rlist}(y),\; 
    \mathrm{rlist}(x)\leftarrow x=\mathrm{NULL}\; \}.
  \end{align*}
  The formula
  $\dform{\ProgP_{llist},\ProgP_{rlist}}{(\mathrm{llist(h_1)} \wedge
    \mathrm{rlist(h_2)})}$ expresses structures where $\mathrm{h_1}$
  is a node on a $\NULL$-terminated list made of
  $\mathrm{left}(\cdot,\cdot)$ pointers and $\mathrm{h_2}$ is a node
  on a $\NULL$-terminated list made of $\mathrm{right}(\cdot,\cdot)$
  pointers. These two lists may be disjoint, like in
  Fig.~\ref{fig:lrlists} (left), or may share nodes, even non-constant
  ones, like in Fig.~\ref{fig:lrlists} (right). There may also be
  other lists in the structure. Notice the difference between
  $\dform{\ProgP_{llist},\ProgP_{rlist}}{(\mathrm{llist(h_1)} \wedge
    \mathrm{rlist(h_2)})}$ and $\dform{\ProgP_{llist} \cup
    \ProgP_{rlist}}{(\mathrm{llist(h_1)} \wedge
    \mathrm{rlist(h_2)})}$. The latter one forbids sharing of
  non-constant nodes, and therefore the structure in
  Fig.~\ref{fig:lrlists} (left) is one of its models while the one in
  Fig.~\ref{fig:lrlists} (right) is not.
\end{example}
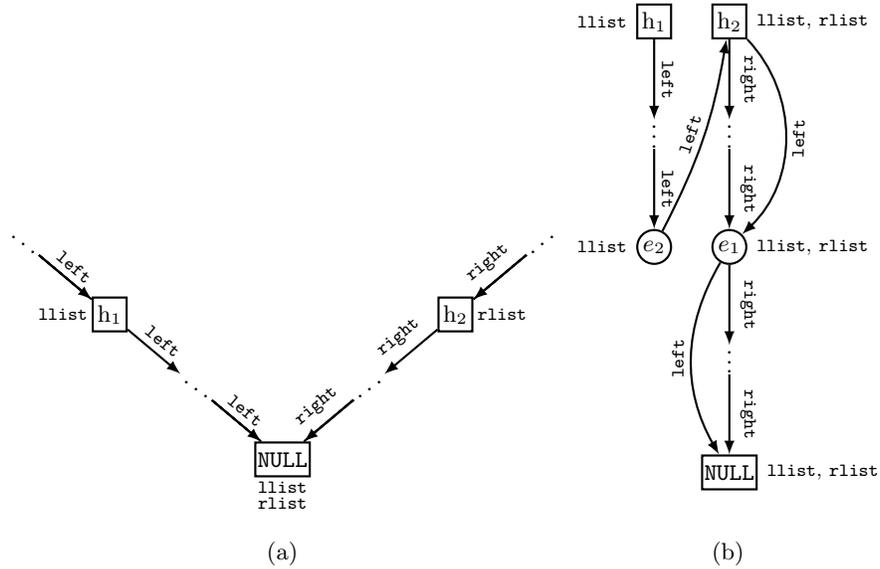
\begin{figure}[htb]
\newcommand{\llist}{\mathord{\mathtt{llist}}}
\newcommand{\rlist}{\mathord{\mathtt{rlist}}}
\vspace{-8ex}
\centering
\subfloat[][]{
\centering
 \begin{tikzpicture}[every node/.style={inner sep = 0mm, minimum size = 0.45cm, shape = circle,      draw}, thick]
 \tikzset{-latex}
 \node (null) [inner sep = 0.3mm, shape = rectangle, draw] {$\NULL$} [grow = up, counterclockwise from=40, sibling angle = 100] { 
  child { node (dots1) [inner sep =0mm, minimum size =0, draw = none, text height = 10pt]  {$\iddots$}  edge from parent[latex-]
  	child {node (h2) [shape = rectangle] {$\mathrm{h_2}$}
  	child { node (dots3) [inner sep =0mm, minimum size =0, draw = none, text height = 10pt]  {$\iddots$}  edge from parent[latex-] 
  	edge from parent node [sloped, above=-3pt,draw = none] {\scriptsize $\rght$}}
  	edge from parent node [sloped, above=-3pt,draw = none] {\scriptsize $\rght$}
  	}
  	edge from parent node [sloped, above=-3pt,draw = none] {\scriptsize $\rght$}
  	}
  child { node (dots2) [inner sep =0mm, minimum size =0, draw = none, text height = 10pt] {$\ddots$} edge from parent [latex-] [counterclockwise from=140]
  	child {node (h1) [shape = rectangle] {$\mathrm{h_1}$}
  		child { node (dots3) [inner sep =0mm, minimum size =0, draw = none, text height = 10pt]  {$\ddots$}  edge from parent[latex-]
  		edge from parent node [sloped, above=-3pt,draw = none] {\scriptsize $\lft$} }
  	  	edge from parent node [sloped, above=-3pt,draw = none] {\scriptsize $\lft$}
  	  	}
  	  	edge from parent node [sloped, above=-3pt,draw = none] {\scriptsize $\lft$}
  	  	}
		};
 \node at (h1.west) [draw = none, left=1pt] {\scriptsize $\llist$};
 \node at (h2.east) [draw = none, right=1pt] {\scriptsize $\rlist$};
 \node at (null.south) [draw = none, below = -6pt] {\scriptsize $\llist$};
 \node at (null.south) [draw = none, below = -0pt] {\scriptsize $\rlist$};
\end{tikzpicture}
}
\subfloat[][]{
 \begin{tikzpicture}[every node/.style={inner sep = 0mm, minimum size = 0.45cm, shape = circle,
      draw}, thick]
 \tikzset{-latex}
 	\node (h2) [shape = rectangle]{$\mathrm{h_2}$}  
 	  	   	child {node (dots2) [inner sep =0mm, minimum size =0, draw = none, text height = 10pt] {$\vdots$} 
 		   		child {node (e1) {$e_1$} 
 		   			child {node (dots3) [inner sep =0mm, minimum size =0, draw = none, text height = 10pt] {$\vdots$} 
 		   				child { node (null)[inner sep = 0.3mm, shape = rectangle, draw] {$\NULL$} 
 		   				edge from parent node [sloped, above=-3pt,draw = none] {\scriptsize $\rght$}
 		   				}
 		   				edge from parent node [sloped, above=-3pt,draw = none] {\scriptsize $\rght$}
 		   				}
 		   				edge from parent node [sloped, above=-3pt,draw = none] {\scriptsize $\rght$}
 		   				}
 		   				edge from parent node [sloped, above=-3pt,draw = none] {\scriptsize $\rght$}
 	 	};
  	\node (h1) [shape = rectangle, left of = h2]{$\mathrm{h_1}$}  
 	  	   	child {node (dots2) [inner sep =0mm, minimum size =0, draw = none, text height = 10pt] {$\vdots$} 
 		   		child {node (e2) {$e_2$} 
 		   				 		   				edge from parent node [sloped, above=-3pt,draw = none] {\scriptsize $\lft$}
 		   				}
 		   				edge from parent node [sloped, above=-3pt,draw = none] {\scriptsize $\lft$}
 	 	};
 	
 	\draw (e2) to [bend right = 9] node [draw = none, sloped, above=-3pt]{\scriptsize $\lft$}  (h2);
 	
 	\draw (h2) to [bend left = 45] node [draw = none, sloped, below=-3pt]{\scriptsize $\lft$}  (e1);
 	
    \draw (e1) to [bend right] node [draw = none, sloped, above=-3pt]{\scriptsize $\lft$}  (null);
      \foreach \name in {null, e1, h2}
          \node at (\name.east) [draw = none, right=3pt] {\scriptsize $\llist$, $\rlist$};
      \foreach \name in {h1, e2}
          \node at (\name.west) [draw = none, left = 3pt] {\scriptsize $\llist$}; 
\end{tikzpicture}
}
\caption{Models of a formula
  $\dform{\ProgP_{llist},\ProgP_{rlist}}{(\mathrm{llist(h_1)} \wedge
    \mathrm{rlist(h_2)})}$ from Ex.~\ref{ex:lrlists}. Lists on the
  left structure are disjoint, with the exception of $\NULL$ node. Lists
  on the right structure share non-constant node $e_1$ and constant
  $\mathrm{h_2}$.  Dots denote arbitrary number of intermediate
  nodes.}
\label{fig:lrlists}
\end{figure}

The formulas written so far did not employ global cardinality
constraints, but, since all of them consist of at most two Datalog
programs, they may be supplemented by such. Consider the example
below. 
\begin{example}\label{ex:halfShared}
  Let us define heaps being two binary trees rooted in
  $\mathrm{r_1}$ and $\mathrm{r_2}$ respectively. We require that the
  number of nodes shared by these two trees is the half of their
  size. A \CTwoDat formula encoding the property is
  $\dform{\Prog{TREE}_1,\Prog{TREE}_2,\Delta}{\phi}$, where
  \begin{align*}
    \Prog{TREE}_1 = \{\; \mathrm{tree_1}(u)&\redl \mathrm{left_1}(u,v)\wedge \mathrm{tree_1}(v)\wedge
    \mathrm{right_1}(u,w)\wedge \mathrm{tree_1}(w),\; \\ \
    \mathrm{tree_1}(u)&\redl u\eq\mathrm{NULL}\; \},
  \end{align*}
  and $\Prog{TREE}_2$ is just $\Prog{TREE}_1$ with every subscript
  $1$ replaced by $2$. The formula $\phi$ is the conjunction of
  \begin{gather*}
    \mathrm{tree_1(r_1)} \wedge \mathrm{tree_2(r_2)} \\
    \forall{u}.\;\mathrm{tree_1}(u) \rightarrow \left(u\eq\mathrm{r_1}
      \vee \exists{v}.\left(\left(\mathrm{left_1}\left(v,u\right) \vee
          \mathrm{right_1}\left(v,u\right)\right)
        \wedge \mathrm{tree_1}\left(v\right)\right)\right) \\
    \forall{u}.\;\mathrm{tree_2}(u) \rightarrow \left(u\eq\mathrm{r_2}
      \vee \exists{v}.\left(\left(\mathrm{left_2}\left(v,u\right) \vee
          \mathrm{right_2}\left(v,u\right)\right)
        \wedge \mathrm{tree_2}\left(v\right)\right)\right) \\
    \forall{u}. \mathrm{shared}(u) \leftrightarrow \mathrm{tree_1}(u)
    \wedge \mathrm{tree_2}(u).
  \end{gather*}
  The global cardinality constraint $\Delta$ is just a single equation
  $\{\#_{\mathrm{tree_1}} + \#_{\mathrm{tree_2}} = 2*\#_{\mathrm{shared}}\}$.

  Existence of both trees is guaranteed by $\Prog{TREE}_1$,
  $\Prog{TREE}_2$ and the first conjunct of $\phi$. Next two conjuncts
  express that $\mathrm{r_1}$ ($\mathrm{r_2}$) must be reached from
  every node labeled by $\mathrm{tree_1}(\cdot)$ (\resp
  $\mathrm{tree_2}(\cdot)$). This effects in that all nodes labeled by
  $\mathrm{tree_1}(\cdot)$ (\resp $\mathrm{tree_2}(\cdot)$) belong to
  the tree rooted in $\mathrm{r_1}$ ($\mathrm{r_2}$). The last
  conjunct of $\phi$ defines auxiliary predicate
  $\mathrm{shared}(\cdot)$ to label exactly the nodes shared by
  both trees. Then, the required cardinality constraint is expressed
  by $\Delta$.
\end{example}

In the examples above we analyzed only shape or quantitative
properties of heaps.  A novel approach to verification of pointer
programs was recently proposed in~\cite{CalvaneseKSVZ13}, where some
properties of heap content are formally specified as description logic
formulas or UML diagrams, and heap shape is defined in a fragment of
separation logic. The last example in this section shows that both
content and shape properties may be expressed in \CTwoDat.

\begin{figure}[htp] \centering{
\includegraphics[scale=1.0]{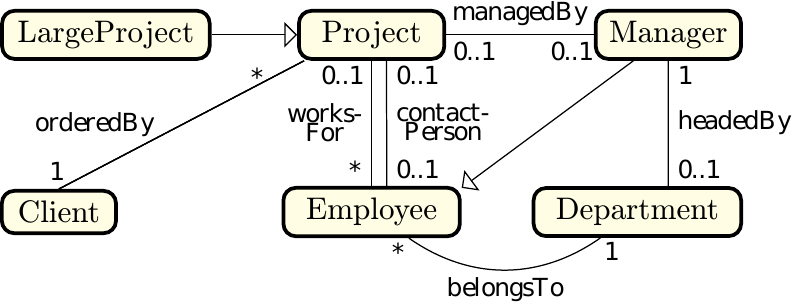}}
\caption{A UML diagram for information system of a company as
  in~\cite{CalvaneseKSVZ13}.}
\label{fig:UMLdiag}
\end{figure}

\begin{example}[Information system of a company, the running example from~\cite{CalvaneseKSVZ13}]\label{ex:uml}
  A software company is divided into departments, has a number of
  employees (some of them are managers), who work for projects (some
  of which are large) and projects are ordered by clients. There are
  certain number restrictions on relations between these entities, as
  specified by UML diagram in Fig.~\ref{fig:UMLdiag}, \eg each
  employee works for at most one project, while each project has an
  arbitrary number of employees working on it. The diagram also
  establishes a subsumption relation between large projects and
  projects (\ie every large project is a project) and similarly for
  managers and employees. Projects, employees, departments and clients
  are stored on $\NULL$-terminated lists on
  $\mathrm{next}(\cdot,\cdot)$ pointers. The information system of the
  company manipulates these lists; it may add and remove their nodes,
  assign managers to departments and projects \etece Every such an
  operation must preserve properties expressed by UML diagram. In this
  example we focus only on defining in \CTwoDat a heap shape and a
  part of the diagram concerning projects, managers and employees. We
  write a formula $\dform{\ProgP_{list}}{\phi}$. First, we have a
  standard list definition, the Datalog program $\ProgP_{list}$ as in
  Example~\ref{ex:simple-list}.
  The formula $\phi$ expresses that
  \begin{enumerate}
    \itemsep0pt
  \item \label{phi:shape}constant nodes $\mathrm{pHd}$ and
    $\mathrm{eHd}$ are heads of two lists; $\mathrm{list(pHd)} \wedge
    \mathrm{list(eHd)} \wedge
    \forall{u}. \left(\neg\mathrm{next}(u,\mathrm{pHd}) \wedge
      \neg\mathrm{next}(u,\mathrm{eHd})\right)$,
  \item \label{phi:listnotrestricted} projects and employees are stored in some lists on heap; \\
    $\forall{u}.\mathrm{project}(u) \vee \mathrm{employee}(u)
    \rightarrow \mathrm{list}(u)$,
  \item all nodes on the list headed in $\mathrm{pHd}$ (\resp
    $\mathrm{eHd}$) are projects(\resp employees);
    $\mathrm{project(pHd)} \wedge \forall{u}.\left(
      \mathrm{project}(u)\rightarrow u\eq\NULL \vee
      \exists{v}.\left(\mathrm{next}(u,v)\wedge\mathrm{project}(v)\right)\right)$
    and a similar formula for $\mathrm{employee}(\cdot)$,
  \item all projects (\resp employees) are on list headed in
    $\mathrm{pHd}$ (\resp $\mathrm{eHd}$);\\
    $\forall{u}.\left( \mathrm{project}(u)\rightarrow u\eq\mathrm{pHd}
      \vee
      \exists{v}.\left(\mathrm{next}(v,u)\wedge\mathrm{project}(v)\right)\right)$
    and a similar formula for $\mathrm{employee}(\cdot)$,
  \item \label{phi:shape-end}projects and employees are disjoint;
    $\forall{u}.\neg\mathrm{project}(u) \vee \neg\mathrm{employee}(u)$
  \item \label{phi:content} each employee has at most one pointer
    $\mathrm{worksFor}(\cdot,\cdot)$ to a project, indicating a
    project that the employee is working on (recall that we are
    writing formula of a logic with functionality restriction);
    $\forall{u}\forall{v}. \mathrm{employee}(u) \wedge
    \mathrm{worksFor}(u,v) \rightarrow \mathrm{project}(v)$,
  \item employees have a Boolean field $\mathrm{is\_manager}(\cdot)$
    marking them as managers;\\ $\forall{u}. \mathrm{is\_manager}(u)
    \rightarrow \mathrm{employee}(u)$,

  \item similarly, projects have a Boolean field $\mathrm{is\_large}(\cdot)$
    marking them as large projects; $\forall{u}. \mathrm{is\_large}(u)
    \rightarrow \mathrm{project}(u)$,

  \item \label{phi:content-end}each project has at most one pointer
    $\mathrm{managedBy}(\cdot,\cdot)$ to an employee being its
    manager; $\forall{u}\forall{v}. \mathrm{project}(u) \wedge
    \mathrm{managedBy}(u,v) \rightarrow \mathrm{is\_manager}(v)$,
  \end{enumerate}
  Conjuncts~\ref{phi:shape}---\ref{phi:shape-end} express heap shape
  properties, \ie that the heap consists of two disjoint lists of projects and
  employees, while conjuncts~\ref{phi:content}---\ref{phi:content-end}
  define properties of heap content, \ie a fragment of the UML
  diagram. One can also include in $\phi$ other properties of the
  information system, not expressed by the UML diagram but encodable in
  the logic, like
\begin{enumerate}
\itemsep0pt
\item[10.] the manager of a project works for the project;\\
$\forall{u}\forall{v}. \mathrm{project}(u) \wedge
    \mathrm{managedBy}(u,v) \rightarrow \mathrm{worksFor}(u,v)$,
\item[11.] at least 10 employees work on each large project\\ 
$\forall{u}. \mathrm{is\_large}(u) \rightarrow \exists^{\geq 10}{v}. 
\left(\mathrm{worksFor}(v,u) \wedge \mathrm{employee}(v) \right)$,
\item[12.] the contact person for a large scale project is a manager;\\ 
$\forall{u}\forall{v}. \mathrm{is\_large}(u) \wedge \mathrm{contactPerson}(u,v) \rightarrow \mathrm{is\_manager}(v)$.
\end{enumerate}
Notice that Conjunct~\ref{phi:listnotrestricted} contains an
unrestricted occurrence of $\mathrm{list}(u)$, thus program
$\Prog{LIST}$ is privileged in $\phi$.
\end{example}

\section{Bounded model checking of pointer programs}\label{sec:bmc}
Imperative pointer programs can naturally be viewed as state
transition systems.  A state stores data structures on heap and values
of program variables in a given program location. Transitions
correspond to program actions.  A transition occurs between two states
if the latter is obtained after successful execution of the
corresponding action in the former state. In general the obtained
transition system is infinite since it models program runs on every
possible initial data structure on heap (\eg a system representing
list reversal program models its execution on every possible finite
list). \emph{Bounded Model Checking of Pointer Programs}~\cite{ChGM05}
aims at discovering presence of $\NULL$-pointer dereferences in
pointer programs, but can also be used for violations of other safety
properties. A counterexample is, roughly, a path from an initial state
to a state where a $\NULL$-pointer dereference occurs. We bound the
length of paths we seek for, but the number of initial states remains
unbounded. Program paths of bounded length are represented by
universal two-variable formulas while admissible initial heap shapes
are described by monadic Datalog programs. Satisfiability of the
obtained formula of two-variable logic with Datalog is then
equivalent to existence of a counterexample. Since logics we consider
are decidable, so is the BMC of pointer programs. The present section
is based on \cite{CharatonikGeorgievaMaier-CSL2005XT} (which is an
extended version of \cite{ChGM05}) with a modified presentation. The
novelty lays in a generalization of the method: apart from
checking for dangling pointers the BMC can now be used to discover
variable aliasing, structure intersection or memory leaks.

\subsection{Syntax of bounded pointer programs}

A \emph{bounded program} $\mathtt{BP}$ (also called a \emph{straight-line
  program}) consists of two parts.  The first one is a \emph{struct
  declaration} specifying types of heap cells (called
\emph{templates} in \cite{ChGM05}). Templates define pointers
(\emph{fields} in \cite{ChGM05}) that start in a given heap cell.  The
second part is a finite sequence of actions specifying possible
program executions.

A \emph{struct declaration} is a finite directed graph with labeled
edges. We call the vertices of this graph \emph{types}, edge
labels are called fields. Formally a struct declaration is a tuple
$\tuple{T, f_1,f_2,\ldots,f_k}$, where $T$ is a set of types and
$f_1,f_2,\ldots,f_k$ are partial functions on $T$.  Every allocated
element of the heap has precisely one type $t\in T$. The meaning of
$f_i(t_1) = t_2$ is that every heap element of type $t_1$
emits a pointer $f_i(\cdot,\cdot)$ to an element of type $t_2$ or to a
special element $\NULL$. Types $t$ are modeled by unary predicates
$t(\cdot)$. Given a structure and a node $e$ of type $t$ in the
structure we call $e$ a \emph{$t$-cell}. Denote by
$\mathrm{struct}(\mathtt{BP})$ the struct declaration of a bounded
pointer program $\mathtt{BP}$.

The set of \emph{actions} $\Act$ is defined by the grammar in 
Figure~\ref{fig:language-actions}, where $t$ is a template, $s$ is a
field, $x$ and $y$ are program variables, $e$ is a program variable or
a constant $\NULL$, and $\gamma$ is an arbitrary \BSTwo formula that
in particular may contain Boolean conditions over program variables
and constants (including $\NULL$) and the equality symbol $\eq$.

\begin{figure}[ht]
\begin{equation*}
\begin{array}{r@{~}c@{~}l@{\qquad}l}
  \Act & \mathrel{::=} & \assume{\gamma}
  & \textit{Skip to next action if condition $\gamma$ is satisfied, fail otherwise.} \\
  & | & y \gts e
  & \textit{Assign the value $e$ to the variable $y$.} \\
  & | & y \gts s(x)
  & \textit{Read the $s$-field of the cell pointed to by $x$ into $y$.} \\
  & | & s(x) \gts e
  & \textit{Write $e$ to the $s$-field of the cell pointed to by $x$.} \\
  & | & \dispose{t}{x}
  & \textit{Deallocate the $t$-cell pointed to by $x$.} \\
  & | & y \gts \new{t}
  & \textit{Allocate a new $t$-cell and assign its address to $y$.}
\end{array}
\end{equation*}
\caption{The action language.}
\label{fig:language-actions}
\end{figure}
Denote by $\mathrm{actions}(\mathtt{BP})$ the sequence of actions of a
bounded pointer program $\mathtt{BP}$.  For an exemplary bounded
pointer program and its struct declaration refer to
Example~\ref{ex:concrete-bmc}.

\subsubsection*{Semantics of bounded programs.} The semantics of actions and
bounded programs is rather self explanatory and can be found in the
thesis~\cite{pwit2014}. We will write $\sem{\str A}{\alpha}{\str B}$
if $\str B$ is obtained by executing an action $\alpha$ in state $\str
A$.  We made only small changes compared to~\cite{ChGM05}, like
introduction of types that allows us to model (un)allocated elements
of heap.

\subsection{The Model Checking Problem for Bounded Programs}
\label{sse:prog:bmc}

Given sets of pre-states and post-states specified by formulas of a
logic with Datalog and a bounded program we want to check if an execution
of the program in some pre-state leads to a post-state. This is
formalized below using \CTwoDat to specify pre- and post-states.

\begin{definition}[Model Checking for Bounded Programs]\label{def:bmc} 
  \ \\ \textbf{Instance}: two \CTwoDat formulas $\varphi =
  \dform{\ProgSeqP,\Delta}\phi$,
  $\varphi' = \dform{\ProgSeqP',\Delta'}{\phi'}$ and a bounded program $\mathtt{BP}$. \\
  \textbf{Question}: does there exist a pre-state $\str A$ and a
  post-state $\str B'$ such that $\str A_{\ProgSeqP} \models \varphi$,
  $\sem{\str A}{\pi}{\str B'}$ and $\str B'_{\ProgSeqP'} \models
  \varphi'$, where $\pi = \mathrm{actions}(\mathtt{BP})$?
\end{definition}

In the above definition formula $\varphi$ is over some vocabularies
$\Sigma_E$ and $\Sigma_I$, and $\varphi'$ is over fresh copies of
these vocabularies, \ie $\Sigma'_E$ and $\Sigma_I'$. Structures $\str
A$ and $\str B$ are $\Sigma_E$-structures, and structure $\str B'$ is
obtained by renaming vocabulary of $\str B$ to its primed version. We
also assume that the vocabulary associated to $\mathtt{BP}$ is
contained in $\Sigma_E$. If the question in the model checking problem
for bounded programs has a positive answer then we say that the
instance $\tuple{\varphi, \varphi', \mathtt{BP}}$ has a solution.  In
this section we assume that all dereference actions of $\mathtt{BP}$
of the form $y \gts s(x)$, $s(x)\gts e$ and $\dispose{t}{x}$ are
prepended with allocation checks $\assume{\alloc{x}}$, where
$\alloc{x}$ is a syntactic shortcut for the formula $\bigvee_{t\in T}
t(x)$ denoting that $x$ is allocated. Note that these checks can be added automatically.

\begin{theorem}[\cite{pwit2014}, Theorem 4.4, Cor. 4.5]\label{th:BMC-reduction-theorem}
  Model checking for bounded programs is polynomially reducible to
  finite satisfiability of \CTwoDat, provided that the total number of
  privileged Datalog programs in $\varphi$ and $\varphi'$ is at most
  $2$. Therefore the problem is \NEXPTIME-complete.
\end{theorem}

The upper bound in the corollary above follows from the observation
that the reduction is polynomial and the satisfiability for the
specification logic is in \NEXPTIME. The lower bound is obtained by a
trivial reduction from satisfiability of the specification logic (take
a formula $\varphi$ and create an instance $\tuple{\varphi,
  \dform{\emptyset}{(x\eq x)}, x \gts x}$).

\subsection{Example analyses}
\label{sec:analyses}
Here we show that model checking for bounded programs can be applied
to some common reasoning tasks, employed \eg in optimizing compilers
(\cite{RepsSaWil07}). One of such analyses is a question whether two
pointer expressions may denote the same heap cell.

\begin{example}[Checking for variable aliasing]\label{ex:aliasing}\  \\
  \textbf{Instance}: A \CTwoDat formula $\varphi$, a bounded program $\mathtt{BP}$ and program variables $x$, $y$. \\
  \textbf{Question}: Does there exist a pre-state $\str A$ such that
  $\str A_{\ProgSeqP} \models \varphi$ and an execution of
  $\mathtt{BP}$ such that $x$ and $y$ reference the same heap cell in
  the post-state? \\
  \textbf{Answer}: Reduction to model checking for bounded
  programs. Formula $\varphi$ and program $\mathtt{BP}$ are already
  defined. Define $\varphi' = \dform{\emptyset}{(x'\eq y')}$.
\end{example}

The question in the above example was about so called
\emph{may-aliasing}. Notice that in our setting we can also answer the
\emph{must-aliasing} question, \ie if two variables refer to the same
heap cell after each execution in each pre-state satisfying a formula;
it is enough to ask if the model checking with the formula
$x'\noteq y'$ does not have solution. Although data structure
traversals encoded by bounded pointer programs are deterministic, may-
and must-aliasing are different problems: think of pre-states
being a singly linked list with two non-$\NULL$ nodes $x$ and
$y$. A bounded pointer program that moves $x$ and $y$ one element
forward $\emph{may}$ produce a post-state where $x$ and $y$ are
aliases, since it happens when these variables are already aliases in
a pre-state. Clearly, they need not be aliases in every
pre-state and the answer to must-aliasing problem is ``no''.

As we have mentioned $\assume{\alloc{x}}$ can be employed to test for
allocation of $x$ before dereferencing or deallocation.  Since testing
for correct dereferencing was one of the main problems solved
in~\cite{ChGM05} we rephrase it as an instance of model checking for
bounded programs. By $\allocp{u}$ we denote the syntactic shortcut for
the primed version of $\alloc{u}$, \ie for the formula $\bigvee_{t\in
  T} t'(u)$.
\begin{example}[Checking for dereference of dangling or $\NULL$
  pointers]\label{ex:dangling}\  \\
  \textbf{Instance}: A \CTwoDat formula $\varphi$, a bounded program $\mathtt{BP}$ and a program variable $x$. \\
  \textbf{Question}: Does there exist a pre-state $\str A$ such that
  $\str A_{\ProgSeqP} \models \varphi$ and an execution of
  $\mathtt{BP}$ such that $x$ is a dangling or $\NULL$ pointer in the
  post-state? \\
  \textbf{Answer}: Reduction to model checking for bounded
  programs. Formula $\varphi$ and program $\mathtt{BP}$ are already
  defined. Define $\varphi' = \dform{\emptyset}{(\neg\allocp{x'})}$.
\end{example}

The next example shows a more realistic extension of the example
above.  It is a slight modification of an example from~\cite{ChGM05}.
A ``real-life'' pointer program gives rise to one or more (possibly
infinitely many) bounded pointer programs obtained by choosing
particular branches in conditional statements, by unwinding of loops
and by inserting $\assume{\alloc{x}}$ actions before dereferencing of
$x$. 

\newcommand{\assgn}[2]{\texttt{{#1}}\(\gts\)\texttt{{#2}}}
\newcommand{\mat}[1]{\begin{math}#1\end{math}}

\begin{example}[Checking for dereference of a dangling or $\NULL$ node
  in a pointer program]\label{ex:concrete-bmc}
Figure~\ref{fi:prog:xmpl_repl:text} shows an example program
$\texttt{PP}_{\cdll}$ taken from \cite{ChGM05}. The struct declaration
of the program is $\tuple{\{\cdllnode\},\nfield,\pfield}$, where both
$\nfield$ and $\pfield$ are $\{\tuple{\cdllnode,\cdllnode}\}$. Upon
start it expects that the variable $\elt$ points to a doubly linked
circular list (realized by $\nfield$- and $\pfield$-pointers).  The
program deallocates the cell pointed to by $\elt$, allocates a new
cell and inserts it in place of the old one (using the temporary
variables $\nelt$ and $\pelt$).

\begin{figure}[htb]
\begin{alltt}
\{ clnode *\nelt; clnode *\pelt, clnode *\elt;  
    \assgn{\nelt}{\nfield(\elt)};
    \assgn{\pelt}{\pfield(\elt)};
    \mat{\dispose{\cdllnode}{\elt}};
    \assgn{\elt}{\mat{\new{\cdllnode}}};
    \assgn{\nfield(\elt)}{\nelt};
    \assgn{\pfield(\elt)}{\pelt};
    \assgn{\nfield(\pelt)}{\elt};
    \assgn{\pfield(\nelt)}{\elt};
\}
\end{alltt}
\caption{Replacing an element in doubly-linked circular list.}
\label{fi:prog:xmpl_repl:text}
\end{figure}

\begin{figure}[htb]

 \begin{minipage}[c]{0.4\textwidth}
\begin{center}
\begin{alltt}
\{ clnode *\nelt; clnode *\pelt,
  clnode *\elt;  
 [0] \mat{\assume{\alloc{\elt}}};
 [1] \assgn{\nelt}{\nfield(\elt)};
 [2] \mat{\assume{\alloc{\elt}}};
 [3] \assgn{\pelt}{\pfield(\elt)};
 [4] \mat{\assume{\alloc{\elt}}};
 [5] \mat{\dispose{\cdllnode}{\elt}};
 [6] \assgn{\elt}{\mat{\new{\cdllnode}}};
 [7] \mat{\assume{\alloc{\elt}}};
 [8] \assgn{\nfield(\elt)}{\nelt};
 [9] \mat{\assume{\alloc{\elt}}};
[10] \assgn{\pfield(\elt)}{\pelt};
[11] \mat{\assume{\alloc{\pelt}}};
[12] \assgn{\nfield(\pelt)}{\elt};
[13] \mat{\assume{\alloc{\nelt}}};
[14] \assgn{\pfield(\nelt)}{\elt};
\}
\end{alltt}

\end{center}
\end{minipage}
\begin{minipage}[c]{.50\textwidth}
\begin{center}
\fbox{
\scalebox{0.5}{
\begin{tikzpicture}
\node (cl)[draw] {$\cdllnode$};
\node (gorny) [above of=cl] {$\nfield$};
\node (dolny) [below of=cl] {$\pfield$};

\draw [->,thick, rounded corners=3pt] (cl.east) -- ++(7pt,0) |- (gorny.south);
\draw [-,thick, rounded corners=3pt] (cl.west) -- ++(-7pt,0) |- (gorny.south);

\draw [->,thick, rounded corners=3pt] (cl.east) -- ++(7pt,0) |- (dolny.north);
\draw [-,thick, rounded corners=3pt] (cl.west) -- ++(-7pt,0) |- (dolny.north);
\end{tikzpicture}
}}
\\
struct declaration
\end{center}
\begin{center}
\fbox{
\begin{tabular}{l}
$\cdll(u) \redl \cdllnode(u), u \eq \elt, \nfield(u,v), \cdll'(v)$.
\\
$\cdll'(u) \redl \cdllnode(u), u \eq \elt$.
\\
$\cdll'(u) \redl \cdllnode(u), u \noteq \elt, \nfield(u,v), \cdll'(v)$.
\end{tabular}
}
\\
Datalog program $\ProgP$
\end{center}
\begin{center}
\fbox{
\begin{tabular}{l}
$\forall u,v \bigl(\pfield(u,v) \iffl \nfield(v,u)\bigr) \andl$
\\
$\forall v \notl \nfield(\NULL,v) \andl$
\\
$\forall v \notl \pfield(\NULL,v)$
\end{tabular}
}
\\
axiom $\phi$
\end{center}
\end{minipage}
\caption{Replacing an element in a doubly linked circular list by a
  new one; the initial condition that in pre-states $\elt$ points to a
  doubly linked circular list is expressed by $\dform{\ProgP}{\left(\phi \andl
  \cdll(\elt)\right)}$.}
\label{fi:prog:xmpl_repl}
\end{figure}

The pointer program supplemented by allocation checks together with
struct declaration for the program and formula $\varphi_{pre} =
\dform{\ProgP}{\left(\phi \andl \cdll(\elt)\right)}$ defining
pre-states are presented in Figure~\ref{fi:prog:xmpl_repl}.  There are
six bounded pointer programs of interest defined by
$\texttt{PP}_{\cdll}$: $\mathtt{BP}_{\text{$[0]$--$[1]$}}$,
$\mathtt{BP}_{\text{$[0]$--$[3]$}}$,
$\mathtt{BP}_{\text{$[0]$--$[6]$}}$,
$\mathtt{BP}_{\text{$[0]$--$[8]$}}$,
$\mathtt{BP}_{\text{$[0]$--$[10]$}}$ and
$\mathtt{BP}_{\text{$[0]$--$[12]$}}$, consisting of line ranges given
in subscripts. For each of these bounded pointer programs we must
check if the dereference occurring after the last line of the bounded
program may fail due to dangling or $\NULL$ pointers. For example, to
be sure that the dereference of $c$ in line $8$ is correct we have to
check that $c$ is allocated after the execution of
$\mathtt{BP}_{\text{$[0]$--$[6]$}}$.  Thus we are interested in the
the following six instances of the model checking problem for bounded
programs.
\begin{gather*}
\tuple{\varphi_{pre}, \mathtt{BP}_{\text{$[0]$--$[1]$}},
  \dform{\emptyset}{(\neg\allocp{c'})}}, \tuple{\varphi_{pre},
  \mathtt{BP}_{\text{$[0]$--$[3]$}}
  ,\dform{\emptyset}{(\neg\allocp{c'})}},\\
 \tuple{\varphi_{pre},
  \mathtt{BP}_{\text{$[0]$--$[6]$}}, 
  \dform{\emptyset}{(\neg\allocp{c'})}}, \tuple{\varphi_{pre},
  \mathtt{BP}_{\text{$[0]$--$[8]$}},
  \dform{\emptyset}{(\neg\allocp{c'})}},\\
 \langle\varphi_{pre},
\mathtt{BP}_{\text{$[0]$--$[10]$}}, \dform{\emptyset}{(\neg\allocp{\pelt'})}\rangle, \tuple{\varphi_{pre}, \mathtt{BP}_{\text{$[0]$--$[12]$}},\dform{\emptyset}{(\neg\allocp{\nelt'})}}.
\end{gather*}
It turns out that $\texttt{PP}_{\cdll}$ is not pointer-safe: the instance
with $\mathtt{BP}_{\text{$[0]$--$[10]$}}$ has a solution. An
analysis of the model of the corresponding formula reveals the
reason.  If $\elt$ points to a circular list of length $1$ then $\pelt
\eq \elt$ after the second action, so $\pelt$ is dangling after
$\dispose{\cdll}{\elt}$.
\end{example}

Problems from  the examples above were already expressible
using the logic from~\cite{ChGM05}. 
The logic was used to specify a pre-state; the specification of post-states
was just a Boolean formula. By contrast, examples below employ more
Datalog programs.

\begin{example}[Checking for structure intersection]\label{ex:intersection}\  \\
  \textbf{Instance}: A \CTwoDat formula $\varphi =
  \dform{\ProgP_1,\ProgP_2}{\phi}$, where $\phi$ together with
  $\ProgP_1$ (\resp $\ProgP_2$) define some linked data structure by
  predicate $shape_1(\cdot)$ (\resp $shape_2(\cdot)$), at most one of
  $\ProgP_1$ and $\ProgP_2$ is privileged in $\varphi$, and a bounded
  program $\mathtt{BP}$. \\ \textbf{Question}: Does there exist a
  pre-state $\str A$ such that $\str A_{\ProgP_1,\ProgP_2} \models
  \varphi$ and an execution of $\mathtt{BP}$ such that structures
  defined by $shape_1(\cdot)$ and $shape_2(\cdot)$ intersect in a
  non-$\NULL$ node in the post-state? \\ \textbf{Answer}: Reduction to
  model checking for bounded programs. Formula $\varphi$ and program
  $\mathtt{BP}$ are already defined. Let $\ProgP'_1$, $\ProgP'_2$ and
  $\phi'$ be $\ProgP_1$ (\resp $\ProgP_2$ and $\phi$) where all
  intensional and extensional are renamed to their primed versions
  (\eg $shape_1(\cdot)$ becomes $shape'_1(\cdot)$). Define the formula
  $\varphi'$ as \[\varphi' = \dform{\ProgP'_1,\ProgP' _2}{\left(\phi'
    \wedge \left(\exists{u}\; shape'_1(u) \wedge shape'_2(u) \wedge
    u\noteq\NULL'\right)\right)}.\]
\end{example}
Since our logic is closed under negation (it is enough to negate the
first order part of a formula) we may also check for negation of the
above properties, \ie for non-dereference of dangling
pointers, variable non-aliasing or structure non-intersection.

In the following example we employ a first-order interpretation of
Datalog programs. A Datalog clause can be seen as a first-order
implication. For a Datalog program $\ProgP$ denote by
$\overline{\ProgP}$ the first order formula being the universally
quantified conjunction of clauses in $\ProgP$. Note that $\str
M_{\ProgP}$ is a model of $\overline{\ProgP}$, but the formula
$\overline{\ProgP}$ may also have other models, \eg the structure
$\str M$ from Figure~\ref{fig:simple}, whose all nodes are labeled by
predicate $\mathrm{list}$ is a model of
$\overline{\ProgP_{\mathrm{list}}}$, but is clearly distinct (\ie
greater) than $\str M_{\ProgP_{\mathrm{list}}}$. We call
$\overline{\ProgP}$ the \emph{the universal closure of $\ProgP$}. By a
simple transformation of formula $\overline{\ProgP}$ one can obtain an
equivalent \FOTwo formula (see Proposition~2.25
in~\cite{pwit2014}). We will use this fact in Example~\ref{ex:leaks}
below.

We say that a bounded pointer program generates a \emph{memory leak}
if it creates a heap node which is allocated but unreachable from any
of program variables.

\begin{example}[Checking for memory leaks]\label{ex:leaks}\  \\
  \textbf{Instance}: A \CTwoDat formula $\varphi_{pre} =
  \dform{\ProgSeqP,\Delta}{\phi_{pre}}$ and a bounded program
  $\mathtt{BP}$. \\ \textbf{Question}: Does there exist a pre-state
  $\str A$, execution of $\mathtt{BP}$ and a node $a\in \str A$ such
  that $\str A_{\ProgSeqP}\models\varphi_{pre}$, the node $a$ is
  either unallocated or allocated and reachable from program variables
  in $\str A$, and $a$ is allocated but unreachable in the post-state?
  \\ \textbf{Answer}: Reduction to model checking for bounded
  programs.  Let $c_a$ be a fresh constant, which will be used to
  denote the above mentioned node $a\in \str A$. Recall that the
  specification logic enforces the bounded-sharing restriction, which
  means, roughly, that only constant nodes may be shared by different
  pointers. Constant $c_a$ is an auxiliary symbol; formally it is a
  constant in the new extensional vocabulary but the node it
  interprets must not be shared unless it also interprets a constant
  from the old vocabulary $\Sigma_E$. To encode the above property we
  use a macro $\mathit{bsr}(u)$ defined as a conjunction of the
  following two formulas.
\begin{gather*}
\bigwedge_{i\in\{1,\ldots,k\}}\bigwedge_{s_1(\cdot,\cdot)\in \Sigma(\ProgP_i)} \forall u_1 \forall u_2\; s_1(u_1,u) \wedge s_1(u_2,u) \wedge u_1\neq
u_2 \rightarrow \bigvee_{c\in\Sigma_E} u = c \\ 
\bigwedge_{i\in\{1,\ldots,k\}}\bigwedge_{s_1(\cdot,\cdot)\in \Sigma(\ProgP_i)}\bigwedge_{s_2(\cdot,\cdot)\in \Sigma(\ProgP_i)\setminus\{s_1\}} \forall u_1 \forall
u_2\; s_1(u_1,u) \wedge s_2(u_2,u) \rightarrow \bigvee_{c\in\Sigma_E} u
= c
\end{gather*}

The instance of the model checking problem will be the tuple
$\tuple{\varphi, \varphi', \mathtt{BP}}$, where $\varphi$ and
$\varphi'$ are described below. Let $\mathrm{struct}(\mathtt{BP}) =
\tuple{T, f_1,f_2,\ldots,f_k}$ and $\mathit{Var}(\mathtt{BP})$ be 
the set of program variables in $\mathtt{BP}$.
 
Let $\mathrm{reach_{pre}}(\cdot,\cdot)$ be a fresh intensional
predicate and $\mathrm{edge}(\cdot,\cdot)$ be a fresh extensional
predicate. Define a Datalog program $\Prog{Q}$ with two clauses
$\mathrm{reach_{pre}}(u) \leftarrow \alloc{u} \wedge u\eq c_a$ and
$\mathrm{reach_{pre}}(u) \leftarrow \alloc{u} \wedge edge(u,v) \wedge
\mathrm{reach_{pre}}(v)$ and a formula $\phi$ as
$\forall{u}\forall{v}\ \mathrm{edge}(u,v) \rightarrow \bigvee^k_{i=1}
f_i(u,v)$.  Intuitively, $\mathrm{reach_{pre}}(u)$ means that $c_a$ is
reachable from $u$.  The following observation will be used to ensure
that $c_a$ is reachable from a program variable $c$ in a pre-state: if
$\str M_{\Prog{Q}}$ is any model of $\dform{\Prog{Q}}{(\phi \wedge
  \mathrm{reach_{pre}}(c))}$ then there is a path from $c$ to $c_a$ in
$\str M_{\Prog{Q}}$ made of edges from
$\{f_1(\cdot,\cdot),\ldots,f_k(\cdot,\cdot)\}$. We are now ready to
define $\varphi$.
\[
\varphi =
\dform{\ProgSeqP,\Prog{Q},\Delta}{\left(\phi_{pre} \wedge \phi \wedge 
  \mathit{bsr}(c_a) \wedge \left(\neg\alloc{c_a} \vee
  \bigvee_{c\in \mathit{Var}(\mathtt{BP})} reach_{pre}(c) \right)\right)}.
\]

Let $\mathrm{reach_{post}}(\cdot,\cdot)$ be a fresh intensional
predicate.  Define a Datalog program $\Prog{R}$ with a clause
$\mathrm{reach_{post}}(u) \leftarrow \alloc{u} \wedge u\eq c_a$ and
clauses $\mathrm{reach_{post}}(u) \leftarrow \alloc{u} \wedge f_i(u,v)
\wedge \mathrm{reach_{post}}(v)$ for every $i\in \{1,\ldots,k\}$.  The
following observation will be used to ensure that $c_a$ is not
reachable from a program variable $c$ in a post-state. Let
$\overline{\Prog{R}}$ be the \FOTwo formula equivalent to the
universal closure of $\Prog{R}$ (it exists by remark a the end of
Section~\ref{sec:monadic-datalog}). If $\str M$ is any model of
$\overline{\Prog{R}} \wedge \neg\mathrm{reach_{post}}(c)$ then there
is no path from $c$ to $c_a$ in $\str M$ made of edges from
$\{f_1(\cdot,\cdot),\ldots,f_k(\cdot,\cdot)\}$. In formula $\varphi'$
defined below we will use $\overline{\Prog{R}}$ instead of $\Prog{R}$
because the Datalog program $\Prog{R}$ enforces bounded sharing on all
edges $\{f_1(\cdot,\cdot),\ldots,f_k(\cdot,\cdot)\}$, while the
first-order formula $\overline{\Prog{R}}$ requires no such a
restriction. This is important since formula $\varphi_{pre}$ describes
pre-states by means of both Datalog program sequence $\ProgSeqP$ and
the first order formula $\phi_{pre}$, and some of
$\{f_1(\cdot,\cdot),\ldots,f_k(\cdot,\cdot)\}$ may appear only in
$\phi_{pre}$ and therefore need not satisfy bounded-sharing for
$\ProgSeqP$. We are now ready to define $\varphi'$.
\[
\varphi' = \dform{\emptyset}{\left(\overline{\Prog{R'}} \wedge
  \allocp{c'_a} \wedge \bigwedge_{c\in \mathit{Var}(\mathtt{BP})} \neg reach_{post}(c')\right)}.
\]

In the formula above, which is just a \CTwo formula, $\Prog{R'}$ is
obtained from $\Prog{R}$ by renaming all its extensional symbols
$\{f_1(\cdot,\cdot),\ldots,f_k(\cdot,\cdot)\}$ to their primed
versions $\{f'_1(\cdot,\cdot),\ldots,f'_k(\cdot,\cdot)\}$ and $c_a$ to
$c'_a$, similarly for $\allocp{\cdot}$. The instance of the model
checking problem is then $\tuple{\varphi, \varphi', \mathtt{BP}}$.

We will now show that program $\mathtt{BP}$ generates memory leak when
run on a state that satisfy $\varphi_{pre}$ if and only if the
instance $\tuple{\varphi, \varphi', \mathtt{BP}}$ has a solution. For
the direct implication assume that $\mathtt{BP}$ runs on state $\str
A$, with $\varphi_{pre}\models \str A$, and generates a memory
leak. Therefore there exists $a\in \str A$ such that either $a$ is
unallocated or allocated and reachable from some constant node $c\in
\str A$ (recall that constant nodes of $\str A$ model variables of
program $\mathtt{BP}$ and $\NULL$). Label node $a$ by a fresh constant
$c_a$.  If $a$ is allocated and reachable from $c$ then take an
arbitrary path from $c$ to $a$ and label its edges by predicate
$\mathrm{edge}(\cdot,\cdot)$. If $a$ is unallocated then we assign no
$\mathrm{edge}(\cdot,\cdot)$ pointers. In both cases the obtained
structure models the formula $\phi$. Let $\str A_{\Prog{Q}}$ be the
least extension of the above modified $\str A$ \wrt $\Prog{Q}$. We
will show that $\str A_{\Prog{Q}}$ satisfies $\varphi$.  Clearly $\str
A_{\Prog{Q}}$ satisfies $\phi_{pre}$ as, by assumption, $\str A\models
\phi_{pre}$. Similarly $\str A_{\Prog{Q}}$ satisfies $\phi$.  Since
$a$ is a node of $\str A$ and $\str A$ satisfies the bsr restriction,
we also infer that $\str A_{\Prog{Q}}$ satisfies
$\mathit{bsr}(c_a)$. If $a$ is unallocated then $\str
A_{\Prog{Q}}\models \neg\alloc{c_a}$. Otherwise, predicate
$\mathrm{reach_{pre}}(\cdot)$ labels a path from some constant $c$ to
$c_a$. Therefore $\str A_{\Prog{Q}}$ satisfies $\varphi$.  Let $\str
B'$ be a structure obtained after execution of $\mathtt{BP}$ on $\str
A$. Node $a$ is allocated, but unreachable from constants. Label $a$
by a fresh constant $c'_a$. Observe that $\str B'\models
\allocp{c'_a}$. Label each node of $\str B'$ that is backward
reachable from $c'_a$ by $\mathrm{reach'_{post}}(\cdot)$. Then $\str
B' \models \overline{\Prog{R}'}$.  Since $a$ is reachable from no
constant $c$, we have $\str B'\models \neg\mathrm{reach_{post}}(c')$
for all $c\in \mathit{Var}(\mathtt{BP})$. Therefore $\str B'\models
\varphi'$. Since $\str A_{\Prog{Q}} \models \varphi$ and $B'\models
\varphi'$ the instance $\tuple{\varphi, \varphi', \mathtt{BP}}$ has a
solution.

Conversely, suppose that $\tuple{\varphi, \varphi', \mathtt{BP}}$ has
a~solution with pre-state $\str A$ and post-state $\str B'$. Since
$\str A\models\varphi$, either the node $c_a$ is unallocated or some
constant $c\in \mathit{Var}(\mathtt{BP})$ is labeled with
$\mathrm{reach_{pre}}(\cdot)$. In the latter case there is a~path from
$c$ to $c_a$ labeled with $\mathrm{edge}(\cdot,\cdot)$, and then the
assumption $\str A\models\phi$ gives that the node $c_a$ is reachable
from a~variable. Therefore $c_a$ is either unallocated or reachable in
the pre-state. Since $\str B'\models\varphi'$, we have that $\str
B'\not\models \mathrm{reach_{post}}(c')$ for all $c\in
\mathit{Var}(\mathtt{BP})$. Observe that $\str B'$ is some model of
program $\Prog{R}'$, so it contains the least model and thus it
contains all atoms $\mathrm{reach_{post}}(u)$ for all $u$ backward
reachable from $c'_a$. Since it does not contain
$\mathrm{reach_{post}}(c')$, the node $c_a$ is not reachable from any
program variable. But it is allocated and thus $\mathtt{BP}$ generates
a~memory leak.
\end{example}

\section{Conclusions, open problems and future work}\label{chap:conclusions}

In this paper we extended the method of bounded model checking of
pointer programs proposed in \cite{ChGM05} by increasing the
expressivity of the logic used for specification of data structures
and properties of programs. We demonstrated expressivity of our logics
on several examples.  The examples provide an evidence of improvement
over the method from~\cite{ChGM05} --- it comes from extended
expressibility of the underlying logics, which gives more
sophisticated description of heaps (as in Examples~\ref{ex:lrlists},
\ref{ex:halfShared} and~\ref{ex:uml}) and new analyses
(Examples~\ref{ex:intersection} and~\ref{ex:leaks}) not expressible in
bounded model checking framework from~\cite{ChGM05}. Notice also that
these analyses can be combined, provided that the number of privileged
Datalog programs in the obtained instance of the model checking
problem is at most $2$.

Our method is based on translation to two-variable logic with counting
quantifiers \CTwo with trees~\cite{ChW13}.  One may ask why we do not
use directly this logic. The most important reason is that in Datalog
it is relatively easy to express common data structures; the semantics
based on least fixed points allows us to control in a simple way
(a)cyclicity of these structures. Trying to express it directly in
\CTwo with trees leads to formulas like our translations, which are
too complicated to be used manually.

Relation of our logics with separation logics, \CTwo and \CTwo with
trees raises a question about possibility of embedding decidable
fragments of separation logics into these logics with counting
quantifiers.

By using unary predicates and the $\assume{\gamma}$ construct we may
model Boolean conditions in (finite unfoldings of) loops and
conditional statements, provided that all data comes from a~finite
domain. We conjecture that a~variant of the logic (the logic \CTwoDat
without privileged Datalog programs, which can be translated to \CTwo
without trees) can be extended to a~logic where data stored on heap
can be accessed by equality tests and then translated to a decidable
logic \CTwo with an equivalence relation~\cite{IPH14}. This would
allow us to extend the analyses expressible in \CTwoDat to cope with
data from infinite domains.

\bibliography{c2bmc}{}

\begin{thebibliography}{10}

\bibitem{AntonopoulosGHKO14}
T.~Antonopoulos, N.~Gorogiannis, C.~Haase, M.~I. Kanovich, and J.~Ouaknine.
\newblock Foundations for decision problems in separation logic with general
  inductive predicates.
\newblock In A.~Muscholl, editor, {\em Foundations of Software Science and
  Computation Structures - 17th International Conference, {FOSSACS} 2014, Held
  as Part of the European Joint Conferences on Theory and Practice of Software,
  {ETAPS} 2014, Grenoble, France, April 5-13, 2014, Proceedings}, volume 8412
  of {\em Lecture Notes in Computer Science}, pages 411--425. Springer, 2014.

\bibitem{ArnoldMSS06}
G.~Arnold, R.~Manevich, M.~Sagiv, and R.~Shaham.
\newblock Combining shape analyses by intersecting abstractions.
\newblock In E.~A. Emerson and K.~S. Namjoshi, editors, {\em Verification,
  Model Checking, and Abstract Interpretation, 7th International Conference,
  {VMCAI} 2006, Charleston, SC, USA, January 8-10, 2006, Proceedings}, volume
  3855 of {\em Lecture Notes in Computer Science}, pages 33--48. Springer,
  2006.

\bibitem{BansalBL09}
K.~Bansal, R.~Brochenin, and {\'{E}}.~Lozes.
\newblock Beyond shapes: Lists with ordered data.
\newblock In L.~de~Alfaro, editor, {\em Foundations of Software Science and
  Computational Structures, 12th International Conference, {FOSSACS} 2009, Held
  as Part of the Joint European Conferences on Theory and Practice of Software,
  {ETAPS} 2009, York, UK, March 22-29, 2009. Proceedings}, volume 5504 of {\em
  Lecture Notes in Computer Science}, pages 425--439. Springer, 2009.

\bibitem{BerdineCalcagnoOHearn04}
J.~Berdine, C.~Calcagno, and P.~O'Hearn.
\newblock A decidable fragment of separation logic.
\newblock In {\em Proc.\ FSTTCS'04}, LNCS 3328, pages 97--109. Springer, 2004.

\bibitem{BiereCCZ99}
A.~Biere, A.~Cimatti, E.~M. Clarke, and Y.~Zhu.
\newblock Symbolic model checking without bdds.
\newblock In R.~Cleaveland, editor, {\em TACAS}, volume 1579 of {\em Lecture
  Notes in Computer Science}, pages 193--207. Springer, 1999.

\bibitem{BrocheninDL08}
R.~Brochenin, S.~Demri, and {\'{E}}.~Lozes.
\newblock On the almighty wand.
\newblock In M.~Kaminski and S.~Martini, editors, {\em Computer Science Logic,
  22nd International Workshop, {CSL} 2008, 17th Annual Conference of the EACSL,
  Bertinoro, Italy, September 16-19, 2008. Proceedings}, volume 5213 of {\em
  Lecture Notes in Computer Science}, pages 323--338. Springer, 2008.

\bibitem{BrocheninDL09}
R.~Brochenin, S.~Demri, and {\'{E}}.~Lozes.
\newblock Reasoning about sequences of memory states.
\newblock {\em Ann. Pure Appl. Logic}, 161(3):305--323, 2009.

\bibitem{BrotherstonFPG14}
J.~Brotherston, C.~Fuhs, J.~A.~N. P{\'{e}}rez, and N.~Gorogiannis.
\newblock A decision procedure for satisfiability in separation logic with
  inductive predicates.
\newblock In Henzinger and Miller \cite{DBLP:conf/csl/2014}, page~25.

\bibitem{CalcagnoGH05}
C.~Calcagno, P.~Gardner, and M.~Hague.
\newblock From separation logic to first-order logic.
\newblock In V.~Sassone, editor, {\em Foundations of Software Science and
  Computational Structures, 8th International Conference, {FOSSACS} 2005, Held
  as Part of the Joint European Conferences on Theory and Practice of Software,
  {ETAPS} 2005, Edinburgh, UK, April 4-8, 2005, Proceedings}, volume 3441 of
  {\em Lecture Notes in Computer Science}, pages 395--409. Springer, 2005.

\bibitem{CalvaneseKSVZ13}
D.~Calvanese, T.~Kotek, M.~Simkus, H.~Veith, and F.~Zuleger.
\newblock Shape and content: Incorporating domain knowledge into shape
  analysis.
\newblock {\em CoRR}, abs/1312.6624, 2013.

\bibitem{ChGM05}
W.~Charatonik, L.~Georgieva, and P.~Maier.
\newblock Bounded model checking of pointer programs.
\newblock In {\em Proceedings of the 19th Annual Conference of the European
  Association for Computer Science Logic ({CSL'05})}, pages 397--412, 2005.

\bibitem{CharatonikGeorgievaMaier-CSL2005XT}
W.~Charatonik, L.~Georgieva, and P.~Maier.
\newblock Bounded model checking of pointer programs.
\newblock Technical Report MPI-I-2005-2-002, Max-Planck-Institut f{\"u}r
  Informatik, 2005.

\bibitem{WChPWit}
W.~Charatonik and P.~Witkowski.
\newblock On the complexity of the {B}ernays-{S}ch{\"o}nfinkel class with
  {D}atalog.
\newblock In C.~Ferm{\"u}ller and A.~Voronkov, editors, {\em Logic for
  Programming, Artificial Intelligence, and Reasoning}, volume 6397 of {\em
  Lecture Notes in Computer Science}, pages 187--201. Springer Berlin /
  Heidelberg, 2010.

\bibitem{ChW13}
W.~Charatonik and P.~Witkowski.
\newblock Two-variable logic with counting and trees.
\newblock In {\em 28th Annual {ACM/IEEE} Symposium on Logic in Computer
  Science, {LICS} 2013, New Orleans, LA, USA, June 25-28, 2013}, pages 73--82.
  {IEEE} Computer Society, 2013.

\bibitem{ClarkeKroeningLerda-TACAS2004}
E.~Clarke, D.~Kroening, and F.~Lerda.
\newblock A tool for checking {ANSI-C} programs.
\newblock In {\em Proc.\ TACAS'04}, LNCS 2988, pages 168--176. Springer, 2004.

\bibitem{CookHOPW11}
B.~Cook, C.~Haase, J.~Ouaknine, M.~J. Parkinson, and J.~Worrell.
\newblock Tractable reasoning in a fragment of separation logic.
\newblock In J.~Katoen and B.~K{\"{o}}nig, editors, {\em {CONCUR} 2011 -
  Concurrency Theory - 22nd International Conference, {CONCUR} 2011, Aachen,
  Germany, September 6-9, 2011. Proceedings}, volume 6901 of {\em Lecture Notes
  in Computer Science}, pages 235--249. Springer, 2011.

\bibitem{Ferrara14}
P.~Ferrara.
\newblock Generic combination of heap and value analyses in abstract
  interpretation.
\newblock In K.~L. McMillan and X.~Rival, editors, {\em Verification, Model
  Checking, and Abstract Interpretation - 15th International Conference,
  {VMCAI} 2014, San Diego, CA, USA, January 19-21, 2014, Proceedings}, volume
  8318 of {\em Lecture Notes in Computer Science}, pages 302--321. Springer,
  2014.

\bibitem{FerraraFJ12}
P.~Ferrara, R.~Fuchs, and U.~Juhasz.
\newblock {TVAL+} : {TVLA} and value analyses together.
\newblock In G.~Eleftherakis, M.~Hinchey, and M.~Holcombe, editors, {\em
  Software Engineering and Formal Methods - 10th International Conference,
  {SEFM} 2012, Thessaloniki, Greece, October 1-5, 2012. Proceedings}, volume
  7504 of {\em Lecture Notes in Computer Science}, pages 63--77. Springer,
  2012.

\bibitem{HabermehlIV06}
P.~Habermehl, R.~Iosif, and T.~Vojnar.
\newblock Automata-based verification of programs with tree updates.
\newblock In H.~Hermanns and J.~Palsberg, editors, {\em Tools and Algorithms
  for the Construction and Analysis of Systems, 12th International Conference,
  {TACAS} 2006 Held as Part of the Joint European Conferences on Theory and
  Practice of Software, {ETAPS} 2006, Vienna, Austria, March 25 - April 2,
  2006, Proceedings}, volume 3920 of {\em Lecture Notes in Computer Science},
  pages 350--364. Springer, 2006.

\bibitem{DBLP:conf/csl/2014}
T.~A. Henzinger and D.~Miller, editors.
\newblock {\em Joint Meeting of the Twenty-Third {EACSL} Annual Conference on
  Computer Science Logic {(CSL)} and the Twenty-Ninth Annual {ACM/IEEE}
  Symposium on Logic in Computer Science (LICS), {CSL-LICS} '14, Vienna,
  Austria, July 14 - 18, 2014}. {ACM}, 2014.

\bibitem{HuthPradhan03}
M.~Huth and S.~Pradhan.
\newblock Consistent partial model checking.
\newblock {\em Electronic Notes in Theoretical Computer Science}, 23, 2003.

\bibitem{ImmermanRabinovichRepsSagivYorsh-CAV04}
N.~Immerman, A.~Rabinovich, T.~Reps, M.~Sagiv, and G.~Yorsh.
\newblock Verification via structure simulation.
\newblock In {\em Proc.\ CAV'04}, LNCS 3114, pages 281--294. Springer, 2004.

\bibitem{IosifRS13}
R.~Iosif, A.~Rogalewicz, and J.~Sim{\'{a}}cek.
\newblock The tree width of separation logic with recursive definitions.
\newblock In M.~P. Bonacina, editor, {\em Automated Deduction - {CADE-24} -
  24th International Conference on Automated Deduction, Lake Placid, NY, USA,
  June 9-14, 2013. Proceedings}, volume 7898 of {\em Lecture Notes in Computer
  Science}, pages 21--38. Springer, 2013.

\bibitem{JacksonVaziri2000}
D.~Jackson and M.~Vaziri.
\newblock Finding bugs with a constraint solver.
\newblock In {\em Proc.\ ISSTA'00}, pages 14--25, 2000.

\bibitem{KlarlundSchwartzbach93}
N.~Klarlund and M.~I. Schwartzbach.
\newblock Graph types.
\newblock In {\em Proc.\ POPL'93}, pages 196--205, 1993.

\bibitem{KotekSVZ14}
T.~Kotek, M.~Simkus, H.~Veith, and F.~Zuleger.
\newblock Towards a description logic for program analysis: Extending {ALCQIO}
  with reachability.
\newblock In M.~Bienvenu, M.~Ortiz, R.~Rosati, and M.~Simkus, editors, {\em
  Informal Proceedings of the 27th International Workshop on Description
  Logics, Vienna, Austria, July 17-20, 2014.}, volume 1193 of {\em {CEUR}
  Workshop Proceedings}, pages 591--594. CEUR-WS.org, 2014.

\bibitem{KuncakLR02}
V.~Kuncak, P.~Lam, and M.~C. Rinard.
\newblock Role analysis.
\newblock In J.~Launchbury and J.~C. Mitchell, editors, {\em Conference Record
  of {POPL} 2002: The 29th {SIGPLAN-SIGACT} Symposium on Principles of
  Programming Languages, Portland, OR, USA, January 16-18, 2002}, pages 17--32.
  {ACM}, 2002.

\bibitem{KuncakR04}
V.~Kuncak and M.~C. Rinard.
\newblock Generalized records and spatial conjunction in role logic.
\newblock In R.~Giacobazzi, editor, {\em Static Analysis, 11th International
  Symposium, {SAS} 2004, Verona, Italy, August 26-28, 2004, Proceedings},
  volume 3148 of {\em Lecture Notes in Computer Science}, pages 361--376.
  Springer, 2004.

\bibitem{LahiriQ08}
S.~K. Lahiri and S.~Qadeer.
\newblock Back to the future: revisiting precise program verification using
  {SMT} solvers.
\newblock In G.~C. Necula and P.~Wadler, editors, {\em Proceedings of the 35th
  {ACM} {SIGPLAN-SIGACT} Symposium on Principles of Programming Languages,
  {POPL} 2008, San Francisco, California, USA, January 7-12, 2008}, pages
  171--182. {ACM}, 2008.

\bibitem{Leino10}
K.~R.~M. Leino.
\newblock Dafny: An automatic program verifier for functional correctness.
\newblock In E.~M. Clarke and A.~Voronkov, editors, {\em Logic for Programming,
  Artificial Intelligence, and Reasoning - 16th International Conference,
  LPAR-16, Dakar, Senegal, April 25-May 1, 2010, Revised Selected Papers},
  volume 6355 of {\em Lecture Notes in Computer Science}, pages 348--370.
  Springer, 2010.

\bibitem{ManevichYRS05}
R.~Manevich, E.~Yahav, G.~Ramalingam, and S.~Sagiv.
\newblock Predicate abstraction and canonical abstraction for singly-linked
  lists.
\newblock In R.~Cousot, editor, {\em Verification, Model Checking, and Abstract
  Interpretation, 6th International Conference, {VMCAI} 2005, Paris, France,
  January 17-19, 2005, Proceedings}, volume 3385 of {\em Lecture Notes in
  Computer Science}, pages 181--198. Springer, 2005.

\bibitem{MollerSchwartzbach01}
A.~M{\o}ller and M.~I. Schwartzbach.
\newblock The pointer assertion logic engine.
\newblock In {\em Proc.\ PLDI'01}, pages 221--231, 2001.

\bibitem{NguyenDQC07}
H.~H. Nguyen, C.~David, S.~Qin, and W.~Chin.
\newblock Automated verification of shape and size properties via separation
  logic.
\newblock In B.~Cook and A.~Podelski, editors, {\em Verification, Model
  Checking, and Abstract Interpretation, 8th International Conference, {VMCAI}
  2007, Nice, France, January 14-16, 2007, Proceedings}, volume 4349 of {\em
  Lecture Notes in Computer Science}, pages 251--266. Springer, 2007.

\bibitem{OHearnRY01}
P.~W. O'Hearn, J.~C. Reynolds, and H.~Yang.
\newblock Local reasoning about programs that alter data structures.
\newblock In L.~Fribourg, editor, {\em Computer Science Logic, 15th
  International Workshop, {CSL} 2001. 10th Annual Conference of the EACSL,
  Paris, France, September 10-13, 2001, Proceedings}, volume 2142 of {\em
  Lecture Notes in Computer Science}, pages 1--19. Springer, 2001.

\bibitem{PekQM14}
E.~Pek, X.~Qiu, and P.~Madhusudan.
\newblock Natural proofs for data structure manipulation in {C} using
  separation logic.
\newblock In M.~F.~P. O'Boyle and K.~Pingali, editors, {\em {ACM} {SIGPLAN}
  Conference on Programming Language Design and Implementation, {PLDI} '14,
  Edinburgh, United Kingdom - June 09 - 11, 2014}, page~46. {ACM}, 2014.

\bibitem{PiskacWZ14}
R.~Piskac, T.~Wies, and D.~Zufferey.
\newblock Automating separation logic with trees and data.
\newblock In A.~Biere and R.~Bloem, editors, {\em Computer Aided Verification -
  26th International Conference, {CAV} 2014, Held as Part of the Vienna Summer
  of Logic, {VSL} 2014, Vienna, Austria, July 18-22, 2014. Proceedings}, volume
  8559 of {\em Lecture Notes in Computer Science}, pages 711--728. Springer,
  2014.

\bibitem{IPH14}
I.~Pratt{-}Hartmann.
\newblock Logics with counting and equivalence.
\newblock In Henzinger and Miller \cite{DBLP:conf/csl/2014}, page~76.

\bibitem{Qiu0SM13}
X.~Qiu, P.~Garg, A.~Stefanescu, and P.~Madhusudan.
\newblock Natural proofs for structure, data, and separation.
\newblock In H.~Boehm and C.~Flanagan, editors, {\em {ACM} {SIGPLAN} Conference
  on Programming Language Design and Implementation, {PLDI} '13, Seattle, WA,
  USA, June 16-19, 2013}, pages 231--242. {ACM}, 2013.

\bibitem{RakamaricBHC07}
Z.~Rakamaric, R.~Bruttomesso, A.~J. Hu, and A.~Cimatti.
\newblock Verifying heap-manipulating programs in an {SMT} framework.
\newblock In K.~S. Namjoshi, T.~Yoneda, T.~Higashino, and Y.~Okamura, editors,
  {\em Automated Technology for Verification and Analysis, 5th International
  Symposium, {ATVA} 2007, Tokyo, Japan, October 22-25, 2007, Proceedings},
  volume 4762 of {\em Lecture Notes in Computer Science}, pages 237--252.
  Springer, 2007.

\bibitem{Rensink04}
A.~Rensink.
\newblock Canonical graph shapes.
\newblock In {\em Proc.\ ESOP'04}, LNCS 2986, pages 401--415. Springer, 2004.

\bibitem{RepsSaWil07}
T.~Reps, M.~Sagiv, and R.~Wilhelm.
\newblock Shape analysis and applications.
\newblock In Y.~N. Srikant and P.~Shankar, editors, {\em The Compiler Design
  Handbook: Optimizations and Machine Code Generation, Second Edition}. CRC
  Press, Inc., Boca Raton, FL, USA, 2nd edition, 2007.

\bibitem{SagivRepsWilhelm02}
M.~Sagiv, T.~Reps, and R.~Wilhelm.
\newblock Parametric shape-analysis problems via 3-valued logic.
\newblock {\em ACM TOPLAS}, 24(2):217--298, 2002.

\bibitem{pwit2014}
P.~Witkowski.
\newblock {\em Complexity of Some Logics Extended with Monadic Datalog
  Programs}.
\newblock PhD thesis, Institute of Computer Science, University of Wroc{\l}aw,
  2014.
\newblock \url{http://www.ii.uni.wroc.pl/~pwit/thesis/thesis.pdf}.

\bibitem{YorshRabinovichSagivMeyerBouajjani}
G.~Yorsh, A.~Rabinovich, M.~Sagiv, A.~Meyer, and A.~Bouajjani.
\newblock A logic of reachable patterns in linked data-structures.
\newblock {\em Journal of Logic and Algebraic Programming}, 73(1-2):111 -- 142,
  2007.
\newblock Foundations of Software Science and Computation Structures 2006
  (FOSSACS 2006).

\bibitem{YorshRepsSagiv04}
G.~Yorsh, T.~Reps, and M.~Sagiv.
\newblock Symbolically computing most-precise abstract operations for shape
  analysis.
\newblock In {\em Proc.\ TACAS'04}, LNCS 2988, pages 530--545. Springer, 2004.

\bibitem{ZeeKR08}
K.~Zee, V.~Kuncak, and M.~C. Rinard.
\newblock Full functional verification of linked data structures.
\newblock In R.~Gupta and S.~P. Amarasinghe, editors, {\em Proceedings of the
  {ACM} {SIGPLAN} 2008 Conference on Programming Language Design and
  Implementation, Tucson, AZ, USA, June 7-13, 2008}, pages 349--361. {ACM},
  2008.

\bibitem{ZeeKR09}
K.~Zee, V.~Kuncak, and M.~C. Rinard.
\newblock An integrated proof language for imperative programs.
\newblock In M.~Hind and A.~Diwan, editors, {\em Proceedings of the 2009 {ACM}
  {SIGPLAN} Conference on Programming Language Design and Implementation,
  {PLDI} 2009, Dublin, Ireland, June 15-21, 2009}, pages 338--351. {ACM}, 2009.

\end{thebibliography}
 
\end{document}